\documentclass[a4paper,twocolumn,superscriptaddress,aps]{revtex4-1}
\usepackage{epsfig}
\usepackage{verbatim}	 % for block comments \begin{comment} \end{comment}
\usepackage{amsmath,amssymb}
\usepackage{calrsfs}
\usepackage{times}
\usepackage{amsmath}
\usepackage{psfrag}
\usepackage{float}
\usepackage{placeins}
\usepackage{color}
\usepackage{bbold}
\usepackage[english]{babel}
\usepackage{graphicx}
\usepackage{braket}
\usepackage{natbib}
\usepackage{hyperref}
\usepackage{lipsum}
\usepackage{multirow}
\usepackage{cleveref}
\usepackage[colorinlistoftodos]{todonotes}

\addto\captionsenglish{}

\DeclareMathAlphabet{\mathcal}{OMS}{cmsy}{m}{n}
\DeclareMathOperator{\sgn}{sgn}

\begin{document}
\title{Emergence of polarized ideological opinions in  multidimensional topic spaces}%non-orthogonal 
%polarization dynamics in a multidimensional ideological space
%Modeling the formation of ideologies due to opinion radicalization
%%main ideas in the title: DATA, model, echo chabers/polarization, radicalization dynamics, networks, social media?

\author{Fabian Baumann}
\thanks{Corresponding author: fabian.olit@gmail.com}
\affiliation{Institut f\"ur Physik, Humboldt-Universit\"at zu Berlin, Newtonstra\ss e 15, 12489 Berlin, Germany}

\author{Philipp Lorenz-Spreen}
\affiliation{Center for Adaptive Rationality, Max Planck Institute for Human Development, Lentzeallee 94, 14195 Berlin, Germany}

\author{Igor M. Sokolov}
\affiliation{Institut f\"ur Physik, Humboldt-Universit\"at zu Berlin, Newtonstra\ss e 15, 12489 Berlin, Germany}
\affiliation{IRIS Adlershof, Humboldt-Universit\"at zu Berlin, Zum Gro\ss en Windkanal 6, 12489, Berlin, Germany}

\author{Michele {Starnini}}
\thanks{Corresponding author: michele.starnini@gmail.com}
\affiliation{ISI  Foundation,  via  Chisola  5,  10126  Torino, Italy}

\begin{abstract}
Opinion polarization is on the rise, causing concerns for the openness of public debates. 
Additionally, extreme opinions on different topics often show significant correlations.
%often can be often correlated with respect to different topics.
The dynamics leading to these polarized ideological opinions pose a challenge: 
How can such correlations emerge, without assuming them a priori in the individual preferences or in a preexisting social structure? 
Here we propose a simple model that {qualitatively} reproduces ideological opinion states found in survey data, even between rather unrelated, but sufficiently controversial, topics.
Inspired by  skew coordinate  systems recently proposed in natural language processing models,
we solidify these intuitions in a formalism 
{of} opinions unfolding in a multidimensional space where topics form a non-orthogonal basis.
{Opinions evolve according to the social interactions among the agents, which are ruled by homophily: two agents sharing similar opinions are more likely to interact.}
The model features phase transition{s} between a global consensus, opinion polarization, and ideological states.
{Interestingly, the ideological phase emerges by relaxing the assumption of an orthogonal basis of the topic space, i.e. if topics thematically overlap.}
{Furthermore, we analytically and numerically show that these transitions are driven by the controversialness of the topics discussed, the more controversial the topics, the more likely are opinion to be correlated.}
Our findings shed light upon the mechanisms driving the emergence of ideology in the formation of opinions.
\end{abstract}

\maketitle

\section{Introduction}

%%change constructive
%%PAR 1: Opinion polarization vs consensus
According to classical opinion dynamics models in which social interactions add constructively to opinion formation, the increasing interaction rates of modern societies would eventually lead to a global consensus, even on controversial issues \cite{degroot1974reaching,RevModPhys.81.591,baronchelli2018emergence}.
This classical prediction has been recently challenged by the empirical observation of opinion polarization, 
i.e. the presence of two well-separated peaks in the opinion distribution.   
Polarization can be found, both offline 
\cite{glaeser2006myths,baldassarri2008partisans}, and in online social media \cite{conover2011political,bail2018exposure,garcia2015ideological}, {where} polarized debates have been observed with respect to several areas and issues, ranging from political orientation~{\cite{adamic2005political, conover2011political,hare2014polarization}}, US and French presidential elections~\cite{Hanna:2013:PAP:2508436.2508438}, to street protests~\cite{Borge-Holthoefer15}. 
Interestingly, polarization seems to burst especially in public discussions evolving around politically and ethically controversial issues such as abortion {\cite{mouw2001culture}} or climate change 
\cite{dimaggio1996have, mccright2011politicization, pew2014political}. {Specifically, in the case of the latter -- climate change -- it has recently been shown, that such polarized non-consensus states hamper the collective resolution of important societal challenges. \cite{Wang17650}.}
%Interestingly, polarization seems to burst especially when issues discussed are controversial, like politics, abortion \MS{More general word for it} or global warming \cite{dimaggio1996have, mccright2011politicization, pew2014political}.
%\cite{abramowitz2008polarization} may be related to the spread of misinformation~\cite{del2016spreading}. 
Different modeling approaches have investigated opinion polarization on single topics as the result of repulsive interactions among agents \cite{Vaz_Martins_2010}, biased assimilation \cite{Dandekar5791}, and information accumulation \cite{shin2010tipping} or social re-inforcement \cite{PhysRevLett.124.048301,banisch2019opinion,mas2013differentiation} mechanisms.

%PAR 2. the problem: opinion correlation
Topics are rarely discussed in isolation. Especially with growing connectedness \cite{hilbert2011world} and increased information flow \cite{lorenz2019accelerating}, the processes of opinion formation take place simultaneously. 
% old version
%{If opinions become polarized, another striking feature can be observed: issue alignment \cite{converse1964nature, baldassarri2008partisans, dellaposta2015liberals}, whose presence implies that individuals are much more likely to have a certain combination of polarized opinions than another combination, a state that can be defined as an ideological opinion state.} 
{For heterogeneous opinion distributions deviating from a global consensus, another striking feature can often be observed: issue alignment \cite{converse1964nature, baldassarri2008partisans, dellaposta2015liberals}, {which has been shown to increase during the recent past \cite{kozlowski2019issue}. The presence of issue alignment} implies that individuals are much more likely to have a certain combination of opinions than others, a state that can be defined as an ideological opinion state.}
For some combinations of topics the alignment is quite intuitive.
For example, opinions with respect to rights of transgender people \cite{jones2020elite} and same-sex couples may be correlated. 
In this case, the majority of individuals would mainly split into two groups, those who deny certain rights to both, transgender people and same-sex couples, and those who support them, 
%Positions such as supporting certain rights for one of the two groups, while denying them to the other group would be rare.
while the mixed positions would be rare.
%{While the positions towards these two gender related issues can clearly be separated in terms of two, well known, ideologies, conservatism and liberalism, other topic combinations, do not offer this direct link. Nevertheless, issue alignment also occur between seemingly unrelated topics.}
While the two gender-related issues can be considered as quite related, 
%and thus the opinion correlation be trivial, other topics are not, as we will see in the following.
in what follows we will show that also opinions on rather unrelated issues might be strongly correlated.
Which underlying mechanism might drive such ideological states to emerge?

%PAR 3: Relevant literature
While considerable efforts have been recently put into measuring and modeling opinion polarization, the phenomenon of issue alignment got much less attention.
This problem has been mainly approached by agent-based modeling within multidimensional opinion spaces, {inspired by Axelrod's seminal work on cultural diversity \cite{axelrod1997dissemination}}.
%In Ref.~\cite{huet2010openness} 
Models based on the concept of a confidence bound illustrated how opinion alignment can result from a dependence between {different} opinion {coordinates} combined with {assimilation and} rejection mechanisms \cite{huet2010openness}, and from assumed correlations between individual and immutable agents' attributes \cite{dellaposta2015liberals,flache2008faultlines}. 
%has been modeled with a tolerance threshold that extends the concept of confidence bounds to two dimensions and adds a rejection mechanism .
%The correlation between individual demographic attributes of agents has been shown to able to generate opinion polarization and issue alignment \cite{flache2008faultlines}.
Other attempts include extensions of both Heider's cognitive balance \cite{heider1946attitudes}{, and argument communication theory \cite{mas2013differentiation}} to multiple dimensions, in well-mixed populations \cite{schweighofer2020,schweighofer2020agent,banisch2019argument}. 
%It has been shown that modular social network structures, with the presence of densely connected clusters, can generate both cultural convergence (consensus) or opinion polarization. 
%\Note{$\rightarrow$ contribution: we don't need network structure, just homophily and co-evolution}

However, all these works assume an a priori, static social network structure (or a well-mixed population) as a substrate for opinion formation, and/or encode issue alignment directly as correlations between individual attributes. 
%include opinion correlations as an ingredient of the modeling framework.
%\Note{instead of this argue that no prescribed rules of interaction or confidence bound is necessary}
On the contrary, social interactions are known to evolve in time \cite{kobayashi2019structured,holme2012temporal}, and such evolution can have a strong impact on the dynamical processes running on top of such time-varying networks, such as opinion formation {and evolutionary games} (see \cite{porter2016dynamical}{\cite{PERC2010109}} for extensive reviews).
This is particularly true for social media platforms, which have been shown to be the major news source for up to $62\%$ of adults in the U.S. \cite{gottfried2016news}. On such platforms the process of opinion formation is continuously shaped by the new information and content shared by users on the platform \cite{holme2014analyzing}.
% 
%\Note{(1) contribution: We generalize this assumption to the concept of homophily. no rejection}\Note{ (2)$\rightarrow$ contribution: we don't need individual attributes}\Note{ (3) not well mixed} \Note{merge sth like "...we aim to minimize the assumption required to obtain polarization and issue alignment" in the beginning.}

%PAR 4. Our solution: emergence of ideological opinion state
%Keywords: connection with non-orthogonal topic space in NLP framework, cosine similarity.
In this paper, we propose a simple model featuring the \emph{emergence} of polarized ideological states from microscopic interactions between individuals, assuming {neither a preexisting social structure}, nor a confidence bound or correlated individual attributes of the agents.
We find that the co-evolution of social interactions and opinions can not only lead to extreme opinions, but can also cause issue alignment. 
Strikingly, such issue alignment emerges also for rather unrelated topics that are sufficiently controversial, due to the reinforcement mechanism mediated by social interactions.
%A small thematic overlap can be non-linearly reinforced through homophilic social interaction, given the opinions of others on one topic influence one's own opinion on another. 
%COMMENT-> This sentence here is not clear, it's too early. what's a thematic overlap?
%describes this process
Our model is based on a minimal set of assumptions.
First, opinions evolve according to the social interactions among the agents, which are ruled by homophily: two agents sharing similar opinions are more likely to interact \cite{PhysRevE.74.056108,PhysRevE.78.016103}.
{This means that the connectivity pattern of the agents is not static but dynamic, the network's evolution is driven by the opinion's dynamics under the assumption of homophilic interactions.
In the same way, opinions evolve according to such social interactions, in a feedback loop leading to a co-evolution of the network's topology and opinion distribution.}
%Opinion evolution can be coupled with the dynamics of the underlying network of social interactions.
%This is particularly true on social media, where the process of opinion formation is continuously shaped by the new information and content shared by users on the platform. 
Second, connected agents sharing similar opinions can mutually reinforce each other's stance. Within the theory of group polarization \cite{isenberg1986group, myers1976group} this happens when individuals, through the exchange of arguments, influence each other in an additive way \cite{vinokur1974effects}.
Third, opinions lay in a multidimensional Euclidean space, spanned by a non-orthogonal basis formed by topics. 
Topics can be controversial and mutually overlapping, i.e. there may exist an intersection of arguments that is valid for several topics.
%Strikingly, issue alignment emerges also for topics with small overlap that are sufficiently controversial, due to the reinforcement mechanism mediated by social interactions.

With these assumptions, our model generates three different scenarios: i) convergence toward a global consensus, ii) polarization of non-correlated opinions, and iii) polarization with issues alignment, i.e. a {polarized} ideological state.
Interestingly, ideology emerges from uncorrelated polarization simply by relaxing the assumption of an {orthogonal basis of the topic space}.
%\tr{I think we should be more careful here. Maybe, along the lines:
%``These three distinct phases---consensus, uncorrelated polarization, and ideology---are not assumed a priori, nor are they driven by global forces, but they can only result from the interactions in a collective of agents. For the connectivity patterns we also do not assume a structure, but let them evolve dynamically, driven by individual activity and pairwise homophily. 
%It is the micro level description of the social system -- summarized by (i) time-varying, homophilic social interactions, (ii) opinions driven by a reinforcement dynamics, and (iii) a non-orthogonal topic space -- that leads to emergence of different macro level configurations, through the co-evolution of the network's topology and the opinion distribution.''}
{These three distinct phases -- consensus, uncorrelated polarization, and ideology -- are neither assumed a priori in the structure of the social interactions, nor are they driven by global forces, but rather emerge from the microscopic interactions among the agents. 
%from the self-organization of the agents, under different conditions. 
It is the micro level description of the social system -- summarized by i) time-varying, homophilic social interactions, ii) opinions driven by a reinforcement dynamics, and iii) a non-orthogonal topic space -- that leads to emergence of different macro level configurations.
%, such as opinions being polarized and correlated, and the social network being segregated according to different opinions.  
}
We analytically and numerically characterize the transitions between these three states, in dependence on the controversialness and overlap of the topics discussed. 
We compare the model's behavior with empirical opinion polls from the American national election surveys (ANES) \cite{anes}. In a pairwise comparison of a broad selection of topics, we can observe several realizations of the scenarios proposed by the model. 
In particular, we found a number of non-trivial cases where opinions are polarized and aligned, but the opinion correlation cannot be simply traced back to the similarity between topics.
%, validating the model's behavior. 
%\tr{(I deleted "validating the model's behavior")}

%emerging phenomena. 

%Our framework is built on a simple one-dimensional model, able to reproduce the polarization dynamics and the presence of echo chambers in polarized debates on Twitter~\cite{PhysRevLett.124.048301}.
Our framework is built on the generalization of a simple one-dimensional model describing polarization dynamics \cite{PhysRevLett.124.048301} to multiple dimensions, assuming the non-orthogonal topic basis. This assumption implies that topics, forming the basis of the space where opinions lay, may not be completely independent but rather can show a certain degree of overlap. 
%By extending this setting to multiple dimensions, ideological states naturally emerge, within the most general setting in {which the underlying  space is not necessarily spanned by an orthogonal topic basis}.
%This assumption implies that topics, forming the basis of the space where opinions lay, may not be completely independent but rather can show a certain degree of overlap. 
%Such non-orthogonality between topics can be interpreted as an inherent thematic overlap of arguments: some arguments for or against an opinion may be valid for both topics.
As suggested by argument exchange theory \cite{burnstein1977persuasive}, a non-vanishing overlap between two topics might arise due to a common set of arguments which simultaneously supports or rejects certain stances on both topics. Thus large overlaps are characteristic for pairs of closely related topics such as our
%This reasoning is most obvious for pairs of very related topics such the above mentioned 
example of rights of transgender people and same-sex couples. 
%Note, however, that also apparently far topics may be related.
%As we will see in the following, 
As we will show, however, also small overlaps critically determine the opinion formation, and hence, ideological opinion states may also emerge for rather unrelated topics.

%By regarding similar parts of life or institutions, some arguments for or against an opinion may be valid for both topics.
%In the above example of issue alignment for opinions regarding rights of transgender people and same-sex marriages, some arguments could be applied to both topics.
%For illustration: In the above example, some normative arguments (e.g., freedom of choice) could be raised and applied to both debates, discussing the rights of transgender people and the ones of same-sex couples.
%\tr{, while other, more specific arguments (e.g., woman's rights, minority rights) could not.}

Interestingly, non-orthogonal {bases} (equivalently, skew coordinate systems) have been recently proposed to solve some well-known problems of classical vector space models for representing text documents \cite{Ando00latentsemantic}. 
Within this framework, documents are represented as vectors in an underlying space, whose basis is formed by the terms used in the documents.
Crucially, if the terms are assumed as orthogonal, similarity measures (such as  cosine similarity) can not precisely describe the relationship between documents, if terms are not independent.  
When the assumption of orthogonality is relaxed, such as in Latent Semantic Indexing or distance metric learning, similarity measures work much better \cite{liu2004learning}.
Our approach follows a similar idea:
if the orthogonality of {topics} is relaxed, i.e. if topics can overlap, the correlation between opinions with respect to different topics can naturally emerge through {the proposed} reinforcement dynamics from social interactions. 

\section{A model of opinion dynamics in a multidimensional topic space}%%%the model
%%%%%%%%%%%
%%Technical note: The idelogical space is an Euclidian space of dimension T with non-orthogonal basis which represent topics. Our coordinates are non-Cartesian (because our basis is not orthogonal). The dot product in our space is defined as a \cdot b^T = \sum_{ij} a_i b_j e_i \cdot e_j, where e_i \cdot e_j = \cos \delta_{ij}. 

Let us consider a system of $N$ agents. Each agent $i$ holds opinions towards $T$ distinct topics, represented by the opinion vector $\mathbf{x}_i=(x_{i}^{(1)}, x_{i}^{(2)},\dots,x_{i}^{(T-1)}, x_{i}^{(T)})$. 
%The opinion vector $\mathbf{x}_i$ represents the position of an agent $i$ in the multidimensional \textit{topic space} $\mathcal{T}$. The opinion vector $\mathbf{x}_i$ can be written as $\mathbf{x}_i = \sum_{v=1}^T x^v_i \mathbf{e}^v$, where $\{x^v_i\}$ are the coordinates of agent $i$ and $\{\mathbf{e}^v\}$ form a basis of the Euclidean space $\mathcal{T}$, representing the topics under consideration.We assume that the considered topics are linearly independent, to form a valid (not necessarily orthogonal) basis of the underlying $T-$dimensional topic space.
In this notation, the component $x_{i}^{(v)} \in [-\infty, +\infty]$  denotes the opinion of agent $i$ towards topic $v$.
%and $T$ is the total number of topics. 
For each topic $v$, the sign of the opinion $x^{(v)}_i$, $\sgn(x^{(v)}_i)$, describes the qualitative stance of agent $i$ towards the topic (i.e., in favor or against the issue), while the absolute value of $x^{(v)}_i$, $|x^{(v)}_i|$, quantifies the strength of his/her opinion, or the conviction, with respect to one of the sides. The opinion vector $\mathbf{x}_i$ represents the position of an agent $i$ in the $T$-dimensional \textit{topic space} $\mathcal{T}$. The opinion vector $\mathbf{x}_i$ can be written as $\mathbf{x}_i = \sum_{v=1}^T x^{(v)}_i \mathbf{e}^{(v)}$, where $\{x^{(v)}_i\}$ are the coordinates of agent $i$ and $\{\mathbf{e}^{(v)}\}$ form a basis of the Euclidean space $\mathcal{T}$, representing the topics under consideration. 
%We assume that the considered topics are linearly independent, to form a valid (not necessarily orthogonal) basis of the underlying $T-$dimensional topic space.
To form the basis in $\mathcal{T}$, $\{\mathbf{e}^{(v)}\}$ have to be assumed linearily independent, but are not necessarily orthogonal.

The opinion vectors of agents evolve in time, i.e.  $\mathbf{x}_i=\mathbf{x}_i(t)$, where we will omit the dependence on $t$ in the following for brevity.
We assume that the evolution of opinions follows a radicalization dynamics, a recently proposed mechanism that reproduces polarization and echo chambers found in empirical social networks \cite{PhysRevLett.124.048301,del2016echo}.
Within this framework, the opinions of an agent are reinforced by interactions with other agents sharing similar views. The mechanism is inspired by the phenomenon of group polarization \cite{isenberg1986group}, by which interactions within a group can drive opinions to become more extreme. 
The social interactions responsible for the opinion dynamics are not static but evolve in time as well \cite{holme2014analyzing,7551567}
%\cite{Barabasi:2005uq}
, forming a time-varying social network that can be represented by a temporal adjacency matrix $A_{ij}(t)$, with $A_{ij}(t)=1$ if agents $j$ and $i$ are connected at time $t$, $A_{ij}(t)=0$ otherwise.
The opinion dynamics is solely driven by interactions among the agents and is described by the following set of $N \times T$ ordinary differential equations,    
\begin{align}
\label{eq:vector}
\dot{x}_i^{(v)} &= -x_i^{(v)} + K\sum_j A_{ij}(t) \tanh\left(\alpha \left[\mathbf{\Phi} \mathbf{x}_j\right]^{(v)}\right)\,,
\end{align}
%\Note{component-wise instead!} \MS{??}
where $K>0$ denotes the social influence strength acting globally among agents -- the larger $K$, the stronger the social influence exerted by the agents on their peers \cite{PhysRevLett.124.048301}. The interpretation of the sigmoidal non-linearity $\tanh(\ldots)$ and the {topic {overlap}} matrix $\mathbf{\Phi}$ will be discussed below.
%\MS{Cannot intro everything here, it's confusing.}
%Furthermore, $\alpha$ and $\mathbf{\Phi}$ denote the controversialness and the \emph{cross topic} matrix, respectively and the non-linearity is applied component-wise to the vector $\mathbf{\Phi x}_j$. 
%Note that we model opinion dynamics as a purely collective, self-organized process without any intrinsic individual preferences. Hence, the opinions of agents lacking social interactions decay towards the neutral state.

According to Eq. \eqref{eq:vector}, the opinion of agent $i$ towards topic $v$, ${x}_i^{(v)}$, evolves depending on the aggregated inputs from his/her neighbors, determined by the temporal adjacency matrix $A_{ij}(t)$. 
The social input of each agent $j$ contributing to the change of $x_i^{(v)}$, $\left[\mathbf{\Phi x}_j\right]^{(v)}$, is smoothed by the influence function $\tanh(\alpha \left[\mathbf{\Phi x}_j\right]^{(v)})$, which tunes the mutual influences that the opinions of different agents exert on each other.   
%The parameter $\alpha$ controls the shape of the function $\tanh(x)$, ensuring that the social influence of extreme opinions is capped, as suggested by experimental findings \cite{jayles2017social}. 
As suggested by experimental findings \cite{jayles2017social}, the social influence
of extreme opinions is capped, and therefore has to be described by a sigmoidal function.
As a particular realization of such function we use $\tanh(x)$, as it was done in the
previous work \cite{PhysRevLett.124.048301}.
%The boundedness of the $\tanh(x)$ function ensures that the social influence of extreme opinions is capped, as suggested by experimental findings \cite{jayles2017social}.
The shape of this function is controlled by the parameter $\alpha$: for small $\alpha$, the social influence of individuals with moderate opinions on other peers is weak, while for large $\alpha$, even moderate agents can exert a strong social influence on others. 
The parameter $\alpha$ can thus be interpreted as the {controversialness} of the topic, which has been shown to be an important factor driving the  emergence of polarization in debates on online social media \cite{Garimella:2018:QCS:3178568.3140565}. 
For the sake of simplicity, we assume $\alpha$ to denote the overall controversy of the discussion around all considered topics, i.e., the same value of $\alpha$ is set for all topics.
The general case of a different controversy for each topic gives rise to further opinion states that can also be found in the empirical data, as shown in the Supplemental Material {\cite{SM}}.
%\tp{In the case of a single topic \cite{PhysRevLett.124.048301}, the parameter $\alpha$ was therefore interpreted as the {controversialness} of the topic, which has been shown to be an important factor driving the  emergence of polarization in debates on online social media \cite{Garimella:2018:QCS:3178568.3140565}. 
%Note, however that for $T>1$ this interpretation does not hold anymore. 
%Hence, $\alpha$ does not correspond to the controversy of a single topic, but denotes the controversy of discussion evolving around the complete set of considered topics. Generalizing, this to multiple $\alpha$ values, one for each topic, gives rise to additional opinion states, as shown in the Supplementary Material.} 
%\Note{Try to show some examples of different $\alpha$ values which could correspond to the empirical cases shown in the SM}

According to Eq. \eqref{eq:vector}, an agent $j$ exerts social influence on a connected agent $i$ with respect to all topics under consideration, 
%where $\left[\mathbf{\Phi x}_j\right]^v$ denotes the social input of agent $j$ about topic $v$
and
%Crucially, we assume that 
the opinion of an agent towards a specific topic is not only influenced by the opinion of others on the same topic but, in general, also about other topics.
%For example, 
This is reflected in the symmetric topic {overlap} matrix $\mathbf{\Phi}$, which encodes the relation between topics.
%i.e. the element ${\Phi}_{v,z}$ quantifies the relation of two topics $v$ and $z$. 
If the element ${\Phi}_{uv}$ is different from zero, the opinions of agents on topic $u$ can influence the opinions of other agents with respect to topic $v$, and {vice versa}. 
%If topics are all independent, the topic matrix reduces to the identity matrix, $\mathbf{\Phi} = \mathbb{1}$. 

\begin{figure}[tbp]
\includegraphics[width=0.7\linewidth]{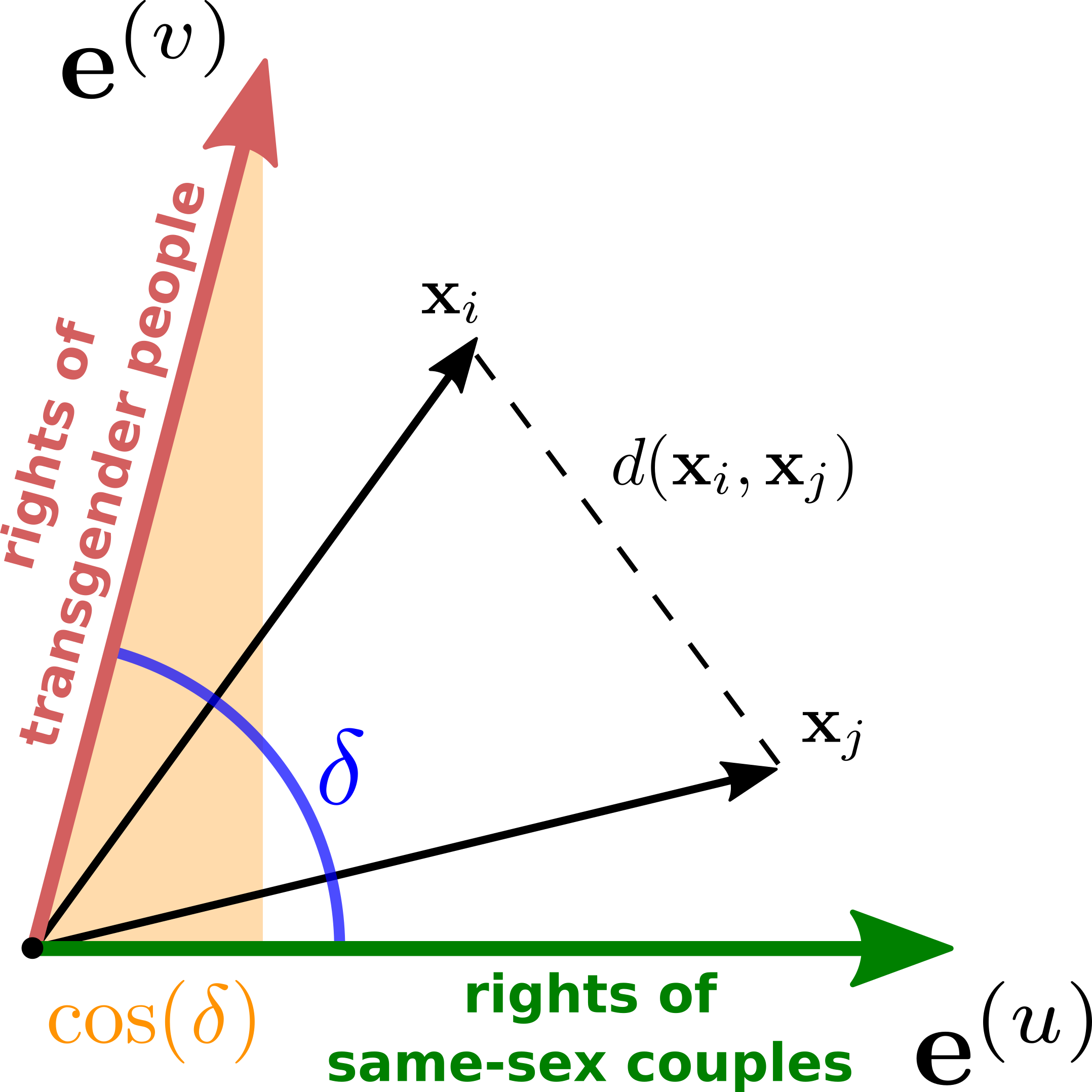}
\caption{\textbf{Illustration of two non-orthogonal topics as basis for the topic space $\mathcal{T}$.} 
For $T=2$, the non-orthogonal, normalized basis is uniquely defined by the angle $\delta$. 
Geometrically, $\cos(\delta)$ quantifies the overlap between basis vectors, interpreted as a topical overlap, here the rights of same-sex couples ($\mathbf{e}^{(u)}$) and transgender people ($\mathbf{e}^{(v)}$). 
The opinion distance between two agents $i$ and $j$, $d(\mathbf{x}_i,\mathbf{x}_j)$, is computed by the scalar product defined in Eq.~\eqref{eq:dot-product}.}
\label{fig:fig0}
\end{figure}

The matrix $\mathbf{\Phi}$ has a geometric interpretation in the latent topic space. 
%Since topics represent the axes of a multidimensional space $\mathcal{T}$, non-vanishing off-diagonal elements of the topic matrix ($\Phi_{v,z}>0$) correspond to relaxing the assumption of an orthogonal basis in the space $\mathcal{T}$. 
The element ${\Phi}_{uv}$ can be interpreted as a scalar product of topics $u$ and $v$, ${\Phi}_{uv} = \mathbf{e}^{(u)} \cdot\, \mathbf{e}^{(v)} = \cos(\delta_{uv})$, where $\delta_{uv}$ represents the angle between topics $u$ and $v$, as shown in Fig.~\ref{fig:fig0} for $T=2$. %\Note{REFER TO FIG0 HERE}. %If topics are orthogonal, one gets $\Phi_{vz}=0$. Note that it always holds $\Phi_{vv}=1$. 
In relation to our introductory example, $\cos(\delta_{uv})$ quantifies the overlap between topic $u$ (rights of transgender people) and $v$ (rights of same-sex couples).
The scalar product between two opinion vectors $\mathbf{x}_i$ and $\mathbf{x}_j$ in the topic space $\mathcal{T}$ spanned by such non-orthogonal topics, is computed as
\begin{equation}\label{eq:dot-product}
 \mathbf{x}_i \cdot \mathbf{x}_j = \mathbf{x}_i^T \mathbf{\Phi} \mathbf{x}_j = \sum_{u,v} {x}_i^{(u)} {x}_j^{(v)} \cos(\delta_{uv})\,,
 \end{equation}
involving the overlap matrix $\mathbf{\Phi}$\,.
%Equation \eqref{eq:dot-product} reveals that the topic matrix $\mathbf{\Phi}$ represents the metric of the underlying topic space. 
%Note that in this Euclidian space, the coordinates $\{x_i^1, x_i^2, \ldots x_i^T\}$ of each agent $i$ are non-Cartesian.
%This representation implies that two non-orthogonal topics are assumed to have some overlap, quantifying their degree of similarity, which vanishes in the case of orthogonal topics. 
%In the example above, $\cos(\delta_{vz})$ quantifies the overlap between topic $v$ (rights of transgender people) and $z$ (rights of same-sex couples). 
Note that it always holds $\Phi_{uu}=1$, so that if all topics are orthogonal, $\Phi_{uv}=0$, the matrix $\mathbf{\Phi}$ reduces to the identity matrix, and Eq.~\eqref{eq:vector} decouples with respect to topics.

The contact patterns among the agents, which sustains the opinion formation, evolves according to the activity driven (AD) model \cite{perra2012activity,PhysRevE.87.062807,Moinet2015,PhysRevLett.112.118702}. 
This gives rise to a temporal network which changes at discrete time intervals. {According to the original AD model, each agent $i$ is characterized by an activity $a_i\in[\varepsilon, 1]$, representing his/her propensity to become active in a given time step. Upon activation, agent $i$ contacts $m$ distinct other agents chosen at random.}
Activities are extracted from a power law distribution $F(a)\sim a^{-\gamma}$, as suggested by empirical findings \cite{perra2012activity, Moinet2015}, such that the parameter set $(\varepsilon, \gamma, m)$ fully defines the basic AD model. 
On top of the basic AD dynamics, we assume that social interactions are ruled by homophily, a well-known empirical feature in both offline \cite{mcpherson2001birds,centola2011experimental} and online \cite{tarbush2012homophily,aiello2012friendship} social networks.
To this end, the probability $p_{ij}$ that an active agent $i$ will contact a peer $j$ is modeled as a decreasing function of the distance between their opinions,
\begin{equation}\label{eq:homophily}
 p_{ij} = \frac{ d( \mathbf{x}_i, \mathbf{x}_j)^{-\beta} }{\sum_j   d( \mathbf{x}_i, \mathbf{x}_j)^{-\beta}},
\end{equation}
where $d( \mathbf{x}_i, \mathbf{x}_j)$ is the usual Euclidean distance between opinion vectors (cf. Fig.~\ref{fig:fig0}) generated by the scalar product defined in Eq.~\eqref{eq:dot-product}, while the exponent $\beta$  controls the power law decay of the connection probability with opinion distance.
As a result of Eq.~\eqref{eq:homophily}, two agents $i$ and $j$ are more likely to interact if they are close in the topic space $\mathcal{T}$, i.e. the distance $d( \mathbf{x}_i, \mathbf{x}_j)$ is small.

{To sum up, at each time step $t$, a different adjacency matrix $A_{ij}(t)$ is generated by the AD dynamics. 
Opinions are subsequently updated on the basis of $A_{ij}(t)$, cf. Eq.~\eqref{eq:vector}.
In the following time step, a new adjacency matrix is generated and opinions are updated accordingly. 
The details of the numerical simulations and the generation of the temporal network are given in Appendix \ref{app:num-sim}.
Therefore, since the generation of $A_{ij}(t)$ depends on the opinions of the agents via homophily, the temporal network co-evolves within the opinion dynamics.}

Upon an interaction between agents $i$ and $j$ (i.e., if $A_{ij}(t)=1$), the opinions of agent $j$ influence all opinions of agent $i$, following the sigmoidal influence function in Eq.~ \eqref{eq:vector}. In the case of orthogonal topics $(\mathbf{\Phi}=\mathbb{1})$ social influence takes place only between opinions on the same topic. 
%\Note{In the light of argument exchange theory this can be interpreted as follows. The arguments exchanged for talking about the respective issues do not \emph{interfere} or \emph{overlap}} 
If the stances of two interacting agents $i$ and $j$ on a topic $u$ are equal, i.e. $\sgn(x_i^{(u)})=\sgn(x_j^{(u)})$, they will increase their current conviction on topic $u$, which is given by the absolute values of the opinion coordinates $\vert x^{(u)}_i\vert$ and $\vert x^{(u)}_j \vert$. On the contrary, for $\sgn(x_i^{(u)})\neq\sgn(x_j^{(u)})$, they will tend to decrease their conviction on that topic and converge towards a consensus. 
%This mechanism corresponds to the one-dimensional radicalization dynamics introduced in \cite{PhysRevLett.124.048301}.
Crucially, for non-orthogonal topics $u$ and $v$, $\cos(\delta_{uv})\neq0$, the opinion with respect to topic $u$ of agent $j$, $x_j^{(u)}$, will influence the opinion of agent $i$ on topic $v$, $x_i^{(v)}$: an argument supporting a topic is logically connected to the other topic.

%\Note{Further small example / half sentence about partially overlapping set of arguments?}
%\MS{OK add it and close paragraph.}
%Therefore, according to Eq. \eqref{eq:vector}, if agents $i$ and $j$ are close in the topic space, they will increase the current conviction of their opinions, given by the norm of the opinion vector $\vert\vert \mathbf{x}_i\vert \vert $, a mechanisms interpreted as a radicalization dynamics. On the contrary, if they are far in the underlying space, they will tend to decrease their opinions' conviction and converge towards the full consensus.\Note{This paragraph needs to be improved. Not clear.}\MS{Why? How?}
%Hence, the radicalization mechanism is generalized to cases where an agent $i$'s conviction towards topic $v$ is only increased if $\sgn(x_i^v)=\sgn(\left[\mathbf{\Phi x}_j\right]^v)$, and viceversa, emphasizing the importance of the underlying metric for the proposed mechanism of social influence. %\Note{Think about better explaing the mechnism for $\mathbf{\Phi}\neq\mathbb{1}$, e.g. write out the sum as $\sgn(x_i^v)=\sgn(\left[\mathbf{\Phi x}_j\right]^v)=\sgn\left(\sum_z\Phi_{vz} x^z_j\right)$} %NO

\section{Emergence of consensus, polarization and ideological phases}%%model behavior 

 %%Summary of results for 1-dim
The model in a one-dimensional space, corresponding to a single topic ($T=1$), has been shown to reproduce empirical data for polarized debates on Twitter,
with respect to polarization of opinions  and segregation of social interactions \cite{PhysRevLett.124.048301}. 
A phase transition between a global consensus
%, in which all agents share similar opinions, 
and polarized state emerged as social influence (tuned by parameter $K$) and the controversialness of the topic discussed (represented by $\alpha$) increased. 
In the following, we explore the impact of multiple topics and their potential overlap within this framework for $T>1$. Following empirical observations, we set the parameters of the basic AD model to $(\epsilon, \gamma, m)=(0.01, 2.1, 10)$ \cite{perra2012activity,PhysRevE.87.062807,Moinet2015,PhysRevLett.112.118702}, and consider a regime of strong social influence and strong homophily, by setting $K=3$ and $\beta = 3$. {In the SM~\cite{SM} we demonstrate that the main findings, presented below, are robust with respect to changes in the AD parameters.}

We investigate the emergence of different opinion states for long times in dependence of $\alpha$ and of the topics overlaps. 
Due to the fluctuations induced by the stochastic interaction dynamics, the states other than consensus are not stable for $t\rightarrow\infty$. 
However, for sufficiently high values of $\beta$ (i.e. homophily), they been shown to be meta-stable \cite{PhysRevLett.124.048301}, numerically indistinguishable from stable states. Therefore, we will refer to them as steady states in the following.
Furthermore, we focus on a regime of fast-switching interactions, i.e. opinions evolve at a slower rate than social interactions. 
This choice is motivated by the assumption that multiple social inputs are necessary to change an agents opinion substantially while attitude change has been shown to be slow, especially in the case of important issues \cite{KROSNICK1988240}.
We therefore choose an integration time step of $dt=0.01$, which corresponds to an effective time-scale separation by a factor of $100$ between the network and the opinion dynamics, see Appendix~\ref{app:num-sim} for details on the numerical simulations.

%\subsection{Different dynamical regimes}

\begin{figure*}[tbp]
\includegraphics[width=\linewidth]{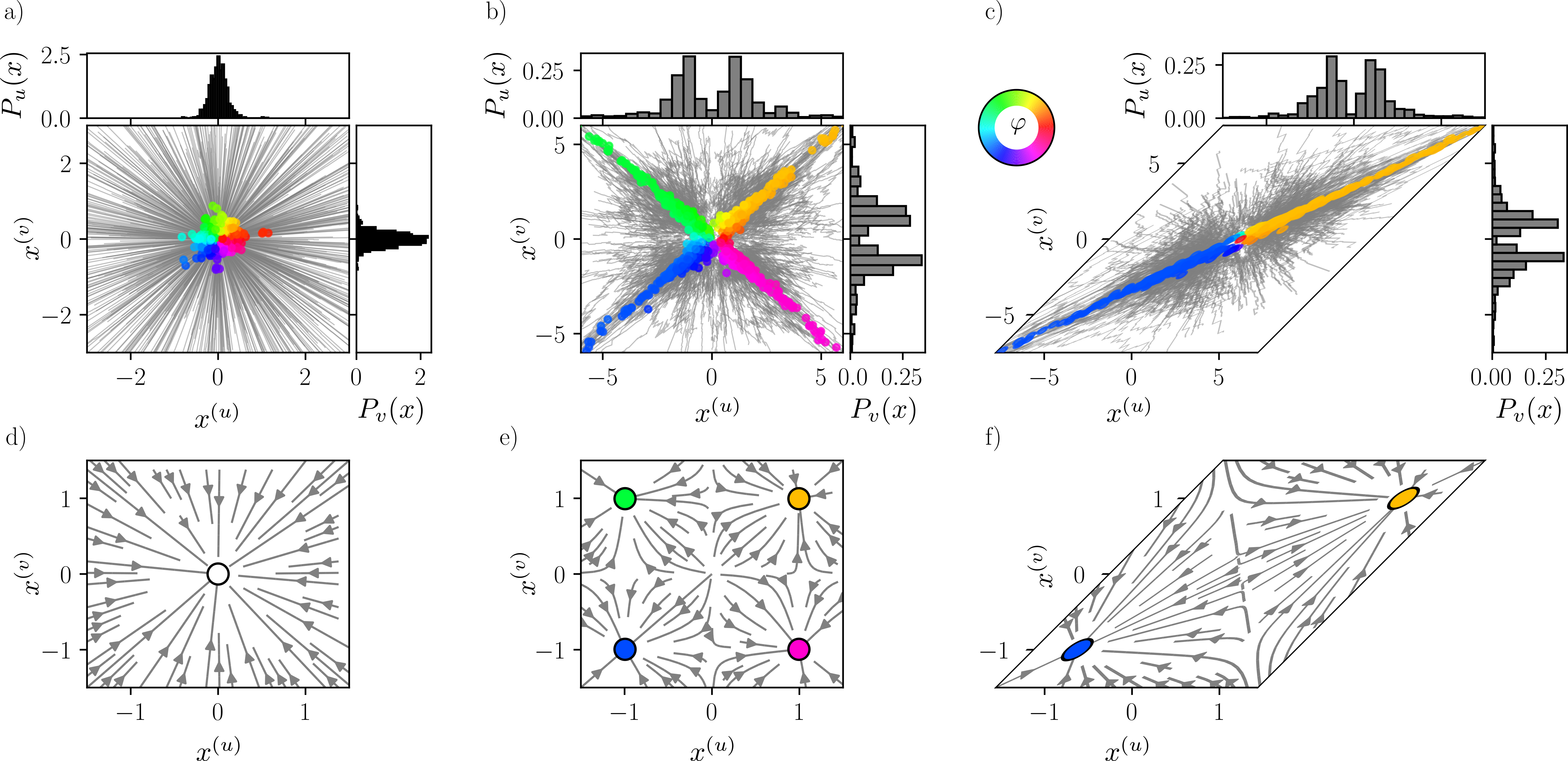}
\caption{\textbf{Temporal evolution of the agents' opinions in a $T=2$ topic space.}
Evolution of opinions from numerical simulations (a)-(c) and  corresponding deterministic dynamics (d)-(f) from mean-field approximation, with identical values of $\alpha$ and $\delta$, cf. Appendix~\ref{app:mf-approxiation} for details. 
The trajectories of the agents' opinions are depicted as grey lines, final opinions are colored according to $\varphi$. 
The system reaches a global consensus if topics are not controversial, for small $\alpha=0.05$ (a), while polarization emerges for controversial topics, for larger $\alpha=3$ (b) and (c). This is indicated by the marginal distributions $P_u(x)$ and $P_v(x)$: the values of the variances are $\sigma^2_u(x) = 0.04$ and $\sigma^2_v(x) = 0.035$ in (a),  $\sigma^2_u(x) = 7.27$ and $\sigma^2_v(x) = 7.17$ in (b), $\sigma^2_u(x) = 11.22$ and $\sigma^2_v(x) = 11.2$ in (c). 
{If topics do not overlap ($\delta=\pi/2$), all combinations of opinion stances appear in consensus (a) and uncorrelated polarized states (b) with low correlation values of $\rho(x^{(u)},x^{(v)})=0.01$ (a) and $\rho(x^{(u)},x^{(v)})=0.024$ (b), respectively.}
If topics overlap ($\delta=\pi/4$ matching the angle between $x$ and $y$ axis in panel (c)), opinions become correlated and ideological states emerge.}
\label{fig:fig1}
\end{figure*}

For the sake of simplicity (and convenient illustrations), in the following we will show the behavior of the model for a system of $N=1000$ agents interacting with respect to two topics $v$ and $u$ ($T=2$). In this case, Eqs.~\eqref{eq:vector} reads
\begin{align}
\dot{x}_{i}^{(u)} &= - x_{i}^{(u)} + K\sum_j A_{ij}(t) \tanh\left(\alpha \left[x_{j}^{(u)}+\cos(\delta)x_{j}^{(v)}\right]\right)
%\label{eq:T2_x} 
\nonumber \\
\dot{x}_{i}^{(v)} &= - x_{i}^{(v)} + K\sum_j A_{ij}(t) \tanh\left(\alpha \left[\cos(\delta)x_{j}^{(u)}+x_{j}^{(v)}\right]\right)%\label{eq:T2_y}
\,,
\label{eq:vector_components}
\end{align}
where  $\mathbf{\Phi}$ is fully defined by a single angle $\delta_{uv}\equiv \delta$, with $\cos(\delta)$ giving the overlap between the two topics considered. {A higher dimensional case with $T=3$ is considered in Sec.~\ref{sec:higher_dimensions}.} 
%The degree of overlap is quantified by $\cos(\delta)$, 
%if there is no overlap, then $\cos(\delta) = 0$ and the cross-topic interaction terms in the influence function $\tanh\!{(x)}$ in Eq.~\eqref{eq:vector_components} vanish.

%The dynamical behavior of the model strongly depends on the form of the cross topic matrix. 

Fig.~\ref{fig:fig1} shows the three dynamical regimes of the model, which strongly depend on the controversialness of topics $\alpha$ and the topic overlap $\cos(\delta)$.
The opinion trajectories of single agents are depicted as grey lines, while their steady state positions are shown as colored dots.
To clarify the visualization, we use polar coordinates $(r,\varphi)$, with $r$ corresponding to the overall conviction of an agent, who is colored according to its opinion, in the polar coordinate $\varphi$. 

If topics are not controversial (i.e. for $\alpha$ small), agents reach a global consensus, as shown in Fig.~\ref{fig:fig1}(a). Starting from normally distributed opinions in the two-dimensional topic space, opinions converge towards the state of vanishing convictions, i.e.  $ \vert \vert \mathbf{x}_i(t\rightarrow\infty) \vert \vert = 0 \, \, \forall i$.
In this regime, the dynamics is dominated by the decay terms $(-x_i^{(u)}, -x_i^{(v)})$ in Eq. \eqref{eq:vector_components}, which mimic the agents' finite opinion memory. 
The fast relaxation toward the global consensus is due to the lack of sufficient social influence from interacting peers. 
This situation is also depicted in the final opinion distributions $P_u(x)$ and $P_v(x)$, plotted on the marginals of  Fig.~\ref{fig:fig1}(a): For both topics, the opinion distribution is peaked around ${x}={0}$.
%%Note that the emergence of consensus is (largely) independent of the topic angle $\delta$, as shown in the following. \Note{we shouldn't write that, cf. Eq.~\ref{eq:stability}}

If topics are controversial -- for larger values of $\alpha$ -- the situation is drastically different, cf. Fig.~\ref{fig:fig1}(b)-(c). 
The social influence among the agents dominates the opinion evolution, destabilizing the global consensus.
%%%'which' what??????
The opinions of agents do not converge but are widely spread and potentially reach convictions much stronger than in the initial configuration.
Note that for polarization to emerge, the presence of homophily is a necessary condition \cite{PhysRevLett.124.048301}. {In the SM~\cite{SM} this is explicitly demonstrated for the model parameters used in Fig.~\ref{fig:fig1}(b), and for increasing levels of homophily in the interval $\beta=[0,3]$. While for vanishing and low $\beta$ non-polarized but radicalized states arise -- as similarly observed in \cite{PhysRevLett.124.048301} in one-dimension -- higher values of $\beta$ change this picture and polarization emerges.}
In this regime, the overlap between topics, encoded by $\cos(\delta)$, crucially determines the dynamics and the possible emergence of ideological states in the system. 
%%%HERE I NEED SENTENCE ABOUT DELTA TO INTRO NEXT PARAGRAPH.

If topics do not overlap, i.e. $\cos(\delta)=0$, the opinions with respect to each topic evolve independently.
That is, the opinion dynamics with respect to each topic decouple, and can be effectively captured by the one-dimensional model of \cite{PhysRevLett.124.048301}.
In this regime of strong social influence, homophily and controversial topics, a polarized state emerges,  as shown in Fig.~\ref{fig:fig1}(b).
In polarized states, the opinion distributions are
bimodal for each topic, as shown on the marginals plots of Fig.~\ref{fig:fig1}(b).
%%%we completely missed description of polarization....
The polarization of opinions with respect to topic $u$ can be quantified by the variance $\sigma^2_u(x)$ of the opinion distribution $P_u(x)$.
A small value of $\sigma^2_u(x)$ implies a consensus-like opinion distribution with respect to topic $u$, while a large $\sigma^2_u(x)$ value indicates polarization. 
The variances $\sigma^2_u(x)$ and $\sigma^2_v(x)$ of the respective marginal distributions are reported in the caption of Fig.~\ref{fig:fig1}. 
For orthogonal topics, all possible combinations of qualitative stances occur, i.e. $[\sgn (x_i^{(u)} ), \sgn (x_i^{(v)})]\in\{(-,+), (+,+),(-,-), (+,-)\}$. 
These four groups, highlighted by different colors in Fig.~\ref{fig:fig1}(b), represent individuals taking all different stances as expected when the two topics are orthogonal. Note that the opinion correlation in both polarized and consensus states is low, as reported in {the caption of Fig.~\ref{fig:fig1}.} %Fig.~\ref{fig:fig1}(a) and Fig.~\ref{fig:fig1}(b).

%for two orthogonal topics. The absence of topic overlap refers to situations in which two topics, albeit discussed simultaneously, do not interfere with respect to exchanged arguments.
%\MS{Add example of orthogonal topics, i.e. social security and religion, whatever.}\Note{Or argument exchange theory as: Arguments for e.g. women's rights will be very different from the ones used in \ldots}
%\Note{Here it is OK to mention an example, in my opinion, as we do not have to hypothesize. It is the observation about opinions directly.}

This situation radically changes if topics overlap ($\cos(\delta)>0$), i.e. they are non-orthogonal in the underlying space.
In this case, according to Eq.~\eqref{eq:vector_components}, the opinions with respect to one topic can influence the opinions with respect to the others, and vice versa.
Fig.~\ref{fig:fig1}(c) shows this situation for $\delta=\pi/4$, i.e. $\cos(\delta)=1/\sqrt{2}$\,. 
At odds with the orthogonal case,
not all combinations of opinion stances are realized in the steady opinion state. 
Instead, the dynamics selects only the opinion states where agents show the same stance on both topics, i.e.   $[\sgn(x_i^{(u)}), \sgn(x_i^{(v)})]\in\{(-,-),(+,+)\}$.
The other stance combinations gradually disappear during approaching the steady state. 
The final opinion distributions $P_u(x)$ and $P_v(x)$ are again bimodal, as shown in the marginal plots of Fig.~\ref{fig:fig1}(c), but the opinions are highly correlated, with the Pearson correlation coefficient $\rho(x^{(u)},x^{(v)}) \simeq 1$.

This state of the system, characterized by opinions which are both polarized, $\sigma^2_u(x), \sigma^2_v(x) \gg 0$, and correlated, $\rho(x^{(u)},x^{(v)}) \gg 0$, is characterized as a \emph{polarized ideological state}. 
In the underlying topic space, this situation translates into a symmetry breaking and consequent dimensionality reduction: The opinion of an agent towards one topic is able to predict his/her opinion towards {the other}. For example, an individual who strongly opposes the idea of same-sex marriage, will also mostly likely argue against transgender people being allowed to use the toilets corresponding to their identified genders.
%\MS{Add another example}
%\Note{OK, example here is good, but should we choose one which we discuss later on specifically in the data-fig or another one?}
%Note that the opinion correlation in both polarized and consensus states is low, as reported in Fig.~\ref{fig:fig1}(a) and Fig.~\ref{fig:fig1}(b), respectively. 
%\tr{Would make it more neutral: ``These qualitatively different scenarios -- consensus, uncorrelated polarization, and ideology -- arise from the micro level description...''}

{It is important to remark that these qualitatively different scenarios -- consensus, uncorrelated polarization, and ideology -- naturally arise from the micro level description of the system, in particular the assumptions of a non-orthogonal topic space $\mathcal{T}$ and social re-inforcement combined with strong homophily. 
More specifically, if topics are not controversial a global consensus is reached, in line with the classical models of opinion averaging \cite{hegselmann2002opinion,doi:10.1142/S0219525900000078,degroot1974reaching}. Instead, if topics are controversial, consensus is not reached and polarization emerges, moving the topic overlaps in the center of attention. Non-overlapping, orthogonal topics yield decoupled opinion dynamics, leading opinions to be separately polarized with respect to each topic. On top of that opinions become correlated, for finite overlaps. Therefore, the three key assumptions of the model -- i) time-varying, homophilic social interactions, ii) opinions driven by a reinforcement dynamics, and iii) a non-orthogonal topic space -- completely determine the dynamics.}

\section{Mean-field approximation}
\begin{figure}[tbp]
\includegraphics[width=\linewidth]{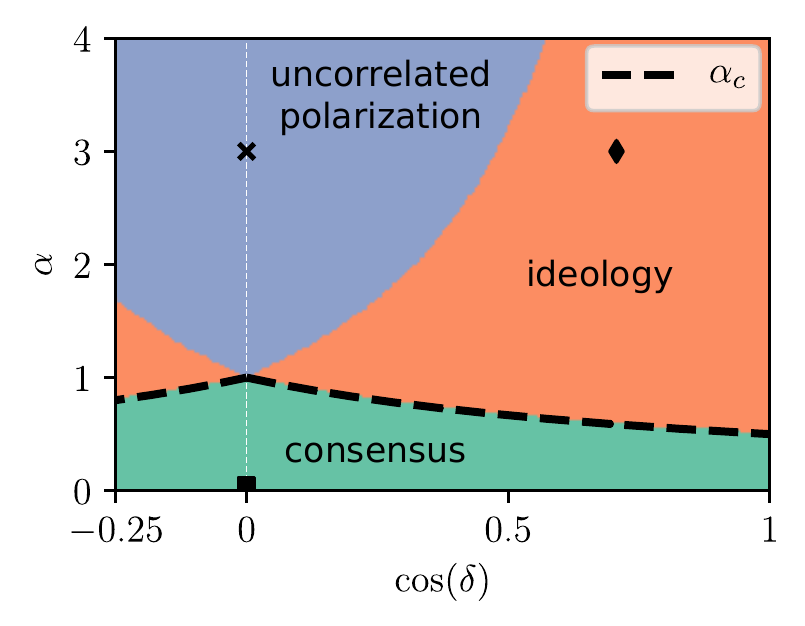}
\caption{\textbf{Stability regions of the mean-field approximation} as a function of the topic overlap $\cos(\delta)$ and controversialness $\alpha$ for ${2}Km\langle a\rangle=1$. 
The different regions in the phase space are colored according to the corresponding states: consensus (green), uncorrelated opinion polarization (blue) and ideological state (orange). 
The black dashed line depicts the critical controversialness $\alpha_c$ separating the regions of consensus and opinion polarization, as given by Eq. \eqref{eq:stability}.
The phase diagram and $\alpha_c$ are symmetric with respect to $\cos(\delta)=0$, i.e. $\delta=\pi/2$, see Appendix~\ref{app:mf-approxiation} for details.
The symbols (square, cross, rhombus) depict the parameter combinations of $\alpha$ and $\cos(\delta)$ used in Fig.~\ref{fig:fig1} and Fig.~\ref{fig:networks}.}
\label{fig:stability}
\end{figure}
The dynamics of the model given by Eq.~\eqref{eq:vector} can, in the thermodynamic limit ($N\rightarrow\infty$) and for strong homophily ($\beta\gg1$), be qualitatively captured within a mean-field approximation, as shown in Appendix~\ref{app:mf-approxiation}. Figures \ref{fig:fig1}(d), (e), and (f) show the attractors of the deterministic, mean-field dynamics for the same values of the parameters $\alpha$ and $\cos(\delta)$ as in Figures \ref{fig:fig1}(a), (b), and (c), respectively. 
The {resulting} dynamics look remarkably similar to the behavior of the full stochastic model.
For low $\alpha$, there is only one stable fixed point, corresponding to the global consensus at  $ \mathbf{x}_i(t\rightarrow\infty) = \mathbf{0} \, \, \forall i$, as shown in Fig.~\ref{fig:fig1}(d). 
As $\alpha$ increases, the consensus is destabilized. If topics are orthogonal, this results in four stable fixed points corresponding to {an uncorrelated} polarized state (Fig.~\ref{fig:fig1}(e)). 
If topics overlap the symmetry is broken and only two stable fixed points emerge, corresponding to the ideological state, depicted in Fig.~\ref{fig:fig1}(f).

\begin{figure*}[tbp]
\includegraphics[width=\linewidth]{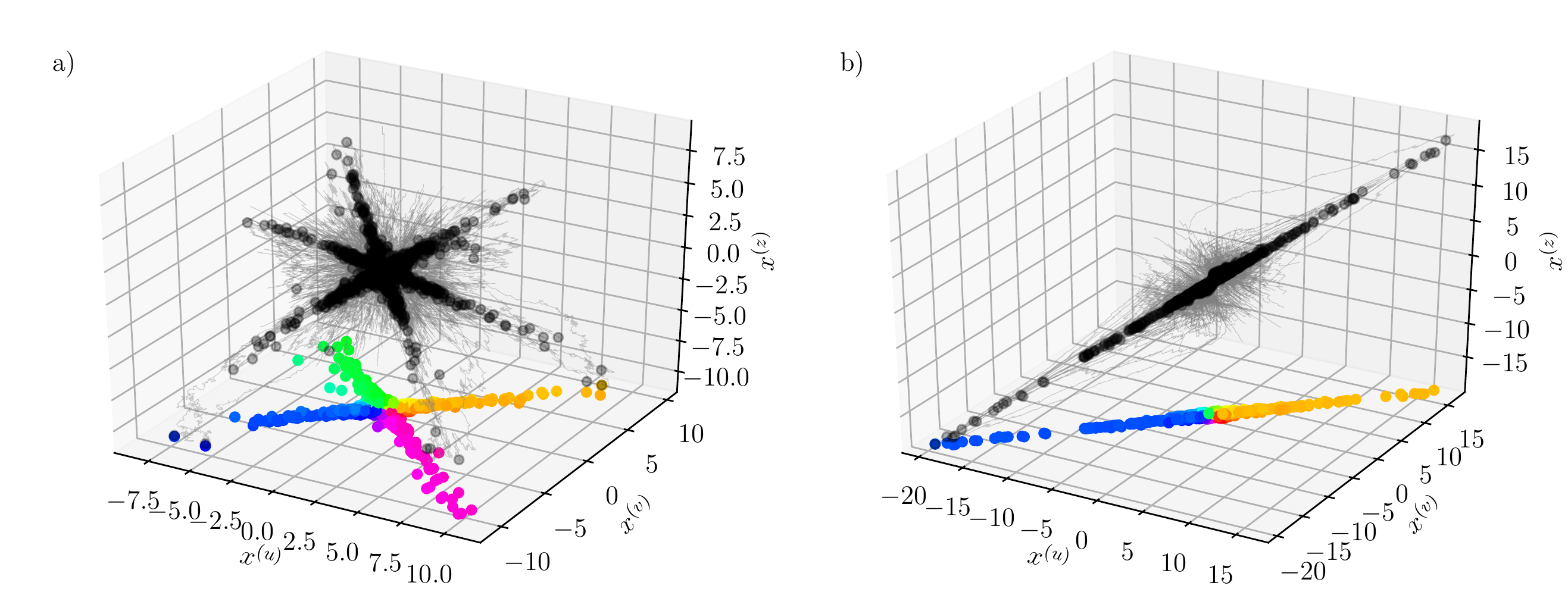}
\caption{{\textbf{Temporal evolution of the agents' opinions in a $T=3$ topic space.}
Evolution of opinions from numerical simulations for strong social influence ($K=3$), controversial topics ($\alpha=3$), and high homophily ($\beta=3)$. 
The grey lines and black dots depict the time evolution of agents' opinions and the steady states, respectively.
In both cases (a) and (b) the topics $u$ and $v$ are orthogonal. 
In panel (a) topic $z$ is also orthogonal to both topics $u$ and $v$. In this case an uncorrelated polarized state emerges, with weak correlations among the three opinions: $\rho(x^{(u)},x^{(v)})=0.15$, $\rho(x^{(u)},x^{(z)})=0.17$ and $\rho(x^{(v)},x^{(z)})=0.11$.
In panel (b) topic $z$ has a finite overlap with both topics $u$ and $v$, i.e. $\cos(\delta_{uz})=\cos(\delta_{vz})=\pi/4$.
In this case, an ideology state emerges: opinions with respect to the three topics are correlated, $\rho(x^{(u)},x^{(v)}) \simeq \rho(x^{(u)},x^{(z)})  \simeq \rho(x^{(v)},x^{(z)})  \simeq 1$. 
Note that for simplicity of illustration the opinion space in panel (b) is depicted using orthogonal axes, although $\delta_{uz}=\delta_{vz}<\pi/2$.}}
\label{fig:non-orthorgonal_xz}
\end{figure*}

Within the mean-field approximation, the transition between a global consensus and polarization can be described analytically. For $T=2$ the stability limits of the consensus phase are determined by the critical controversialness, $\alpha_c$, as
\begin{align}
\label{eq:stability}
    \alpha_c=\frac{1}{{2}K{m}\langle a\rangle [1+\cos(\delta)]}\,,
\end{align}
{which is depicted in Fig.~\ref{fig:stability} as black dashed line.} It depends inversely on the product of social influence strength $K$, {the number of agents contacted by an active agent $m$}, the average activity $\langle a \rangle$, and a factor $[1+\cos(\delta)]$ accounting for the overlap of the two topics.
%The stability of the polarized states can be determined numerically, to distinguish between polarization of non-correlated opinions and the ideological phase, see Methods section for details.
The different regimes of polarization, i.e. polarization of non-correlated opinions and the ideological phase can be distinguished numerically, see Appendix~\ref{app:mf-approxiation} for details.

{Eq. \eqref{eq:stability} thus provides insights for the prediction of the emergence of the ideological phase with respect to the discussion of two topics. 
For instance, between two pairs of topics with similar, small thematic overlap (small $\cos(\delta)$), ideology is expected to emerge more likely for the more controversial pair (larger $\alpha$).
Similarly, Eq. \eqref{eq:stability} shows that the critical controversialness $\alpha_c$ needed for the emergence of the ideological phase is inversely related to the social interaction rate, represented by $m \langle a \rangle$. 
This implies that within contexts where social interactions happen more frequently, such as online social media, even less controversial topics can lead to the emergence of ideology.}

%phase space
Figure~\ref{fig:stability} shows the stability regions in the $\alpha$-$\cos(\delta)$ plane, colored according to the corresponding phases,  consensus (green), polarization of uncorrelated opinions (blue), and ideology (red).
%first, difference overlap/no overlap
Note that the phase diagram is symmetric with respect to the line of vanishing overlaps $\cos(\delta)= 0$ (orthogonal topics). For this case, no ideological states emerge. 
{Note, however, for $\alpha=1$ the ideological phase extends until $\cos(\delta)=0$, giving rise to a triple point, where all three phases, i.e. consensus, uncorrelated polarization, and ideology, coincide. 
This suggests that, closely around $\alpha=1$, ideological states may emerge for already infinitely small overlaps, as we show in App.~\ref{app:mf-approxiation}. For growing overlaps the region of stability for ideological states (orange) widens. Hence, the larger the overlap between topics (the larger the value of $\cos(\delta)$), the smaller is the critical controversialness $\alpha_c$ necessary to de-stabilize consensus and promote ideology, as given by Equation~\eqref{eq:stability} (plotted as a dashed line in Fig~\ref{fig:stability}).}
% OLD AND CHANGED
%By contrast, for finite overlaps, $\cos(\delta)>0$, i.e. non-orthogonal topics, ideological states emerge and their region of stability (red region) widens as the topics' overlap, $\cos(\delta)$, increases. If topics are sufficiently controversial, i.e. for  $\alpha>\alpha_c$, as given by Equation~\eqref{eq:stability} (plotted as a dashed line in Fig~\ref{fig:stability}), consensus is de-stabilized and polarization emerges. The larger the overlap between topics (the larger the value of $\cos(\delta)$), the smaller is the critical controversialness $\alpha_c$ necessary to de-stabilize consensus and promote polarization.

{The phase transition from uncorrelated polarization to ideological states is also driven by the overlap between the topics, $\cos(\delta)$, as shown in Fig.~\ref{fig:stability}. 
The transition is sharp with respect to this parameter: for a certain value of the topic overlap, the final configuration of the agents changes from uncorrelated polarization to the ideological phase. 
In the same way, the phase transition between global consensus and ideology is highly non-linear as a function of the controversialness of the topics $\alpha$, as indicated by Equation~\eqref{eq:stability}.}

% In fact, the region of stable ideological states lies between the full consensus (purple) and uncorrelated polarized (yellow) phase, which suggests the following. While generally, large controversialness gives rise to polarization by de-stabilizing the consensus phase, increasing $\alpha$ values can also trigger a transition from ideology states to uncorrelated polarized states for $\cos(\delta)>0$. Hence, the model suggests that large controversialness does not only destroy an established consensus, but might also decorrelate opinion states in the polarized regime for even larger values of $\alpha$.
%As suggested by Fig.~\ref{fig:stability} the emergence of the ideological phase is intimately connected to situations of $\delta<\pi/2$. 
%In such cases the ideological phase (cyan region) separates the phases of full consensus and uncorrelated opinion polarization, respectively. 
%This suggest, that for $\delta<\pi/2$, increasing $\alpha$ does not only lead to the de-stabilization of the global consensus. A further increase, of the topics' controversialness also  prohibits the establishment of ideological states. Hence, for too large values of the system is not able to maintain coherent relations between the agents' opinion. 
%\Note{Formulate much better, but important: come back to this in the conclusions} 

\section{Higher dimensional case}\label{sec:higher_dimensions}
{Up to this point, we only analyzed the simplest case of two dimensions. 
In this section, we study the behavior of the model in a higher dimensional case.
While for two topics $u$ and $v$ ($T=2$) all potential pairwise topic relations are encoded in a single parameter, their mutual overlap $\cos(\delta_{uv})$, the number of pairwise angles grows quadratically, as $T(T-1)/2$, with increasing dimensions $T$.
For this reason, let us consider only three topics $u$, $v$, and $z$.
This scenario can be effectively described by the three topic overlaps, namely $\cos(\delta_{uv})$, $\cos(\delta_{vz})$, and $\cos(\delta_{uz})$, whose interplay we will explore in the following.
In particular, if two topics are orthogonal, i.e. $\cos(\delta_{uv})=0$, which is the effect of the third topic $z$ on the emergence of correlations between topics $u$ and $v$?}

{As before, let us focus on a regime of strong social influence ($K=3$) and high homophily ($\beta=3$). 
Since we are interested in polarized (correlated or not) states, let us also assume that topics are very controversial ($\alpha=3$). 
If topic $z$ does not overlap with the other two topics ($\cos(\delta_{vz})=\cos(\delta_{uz})=0$), an uncorrelated polarized state emerges, confirming the picture observed for the two-dimensional case. 
In three dimensions, the uncorrelated state is depicted in Fig.~\ref{fig:non-orthorgonal_xz}(a), with final opinions shown as black dots. 
This behavior is analogous to the one shown in Fig.~\ref{fig:fig1}(b), which becomes even clearer when considering the projection of the three-dimensional opinion state on the two-dimensional $(u,v)$-plane. 
Here each dot corresponds to one agent's opinion, color coded according to the opinion angle $\varphi_i$ with respect to topics $u$ and $v$. The projection reveals the same pattern as observed in Fig.~\ref{fig:fig1}(b). The overall pairwise opinion correlations are very low, as reported in the caption of Fig.~\ref{fig:non-orthorgonal_xz}.}

\begin{figure*}[tbp]
\includegraphics[width=\linewidth]{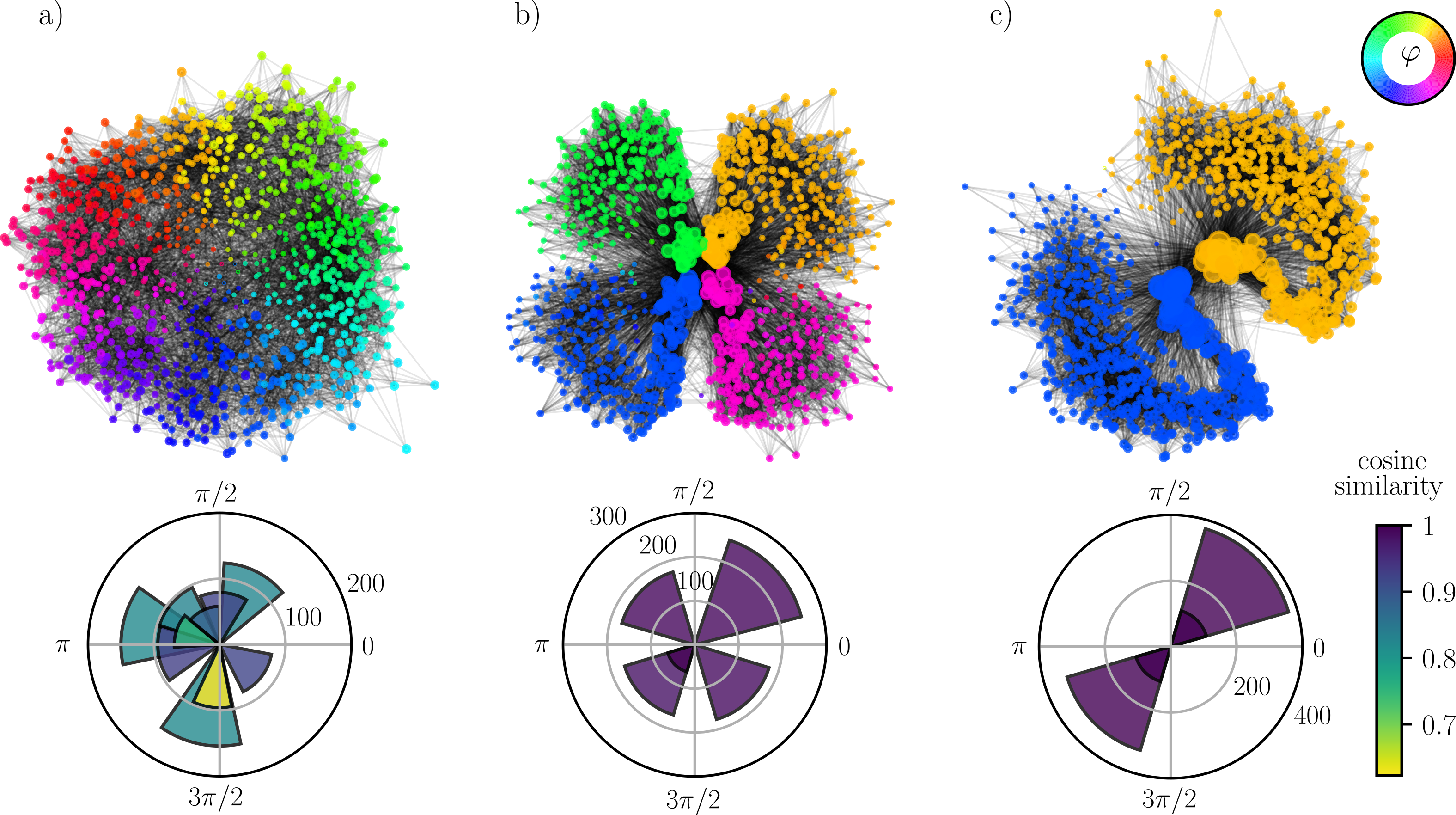}
\caption{\textbf{Community structure of the social networks.}
Visualization of the social networks aggregated over the last 70 time steps (top) and  corresponding community detection (bottom) for three different dynamical regimes: (approaching) consensus (a), uncorrelated polarized state (b) and ideological state (c). 
The model parameters were set as in Fig.~\ref{fig:fig1}(a)-(c), i.e. $\alpha=0.05$, $\delta=\pi/2$ (a), {$\alpha=3$}, $\delta=\pi/2$ (b), {$\alpha=3$}, $\delta=\pi/4$ (c). 
In the network illustrations each node is colored according its opinion angle $\varphi$, and its size is proportional to its conviction $r$. 
Communities are represented in the polar bar plot below each network.  
Each community is represented by a bar: the radius represents the size, color and width correspond to the average cosine similarity between all pairs of agents within the community.
The orientation represents the average opinion angle $\langle\varphi\rangle$ of all agents within the community. Communities containing less than $5\%$ of the total number of nodes are not shown.}
\label{fig:networks} 
\end{figure*}

{Let us now consider the case of the third topic $z$ having finite overlaps with the other two topics, which we assume to be identical for the sake of simplicity, i.e. $\cos(\delta_{uz})=\cos(\delta_{vz})>0$.
This leads to a polarized ideological state, shown in Fig.~\ref{fig:non-orthorgonal_xz}(b),
where a high opinion correlation with respect to topics $u$ and $v$ emerges ($\rho_{uv}\simeq 1$), although topics $u$ and $v$ remain orthogonal. 
The agents' opinions, projected in $(u,v)$-plane, are distributed precisely as in the two dimensional case, cf. Fig.~\ref{fig:fig1}(c). This indicates that an ideological state may emerge even regarding topics entirely unrelated (i.e. orthogonal in this framework), as $u$ and $v$, if the topic space is expanded to higher dimensions and other, related topics (such as topic $z$) are taken into account.}

%\rev{This higher dimensional case has a few implications. Note, that the very definition of the relevant topics in the public discussion is indeed difficult. Within the proposed framework, this means that the number of dimensions is not fixed a priori, such that our results provide a possible explanation for the emergence of opinion correlations between two completely unrelated topics. Namely, that correlations between two topics might be due to the presence of a relevant third topic, related to the previous two, that needs to be included in the analysis. Such confounding topics may well be present, although they are not covered by the empirical data set. Hence, our framework may inform the search for such ``hidden dimensions''.}

{This higher dimensional case has a few implications.
Note, that the very definition of the relevant topics in the public discussion is difficult. Within the proposed framework, this means that the number of dimensions is not {known} a priori, such that our
results provide a possible explanation for the emergence of opinion correlations between two completely unrelated topics. 
Namely, that correlations between two topics might be due to the presence of a relevant third topic, related to the previous two, that needs to be included in the analysis. Such confounding topics may well be present, although not covered by the empirical data set. 
Hence, our framework may {suggest the search} for such hidden dimensions.}

%Those mediators may correspond to e.g. an influential public figure expressing strong opinions on both such topics, to certain views on religion, or to mass media with a mediating influence.

%Therefore, we consider a topic space of two seemingly unrelated topics $X$ and $Y$,  e.g. ("\emph{Transgender bathroom}") and ("\emph{wall with Mexico}"), that is complemented by a third topic $Z$. Assuming pairwise orthogonal topics, $X$, $Y$ and $Z$, 

%The emerging states similar to the ones emerging for two dimensions and a small overlap. \tp{go on here!}

%There are indeed several other factors that can contribute to the emergence of opinion correlation between topics $X$ and $Y$ (such as views on religion or media influence).
%Other important societal features, such as media influence, may mediate influence between a specific pair of topics. 
%However, as the number of potential mediators is very large (and many of them are not observable), it is impossible to include all of them within a tractable modeling framework. 
%Hence, instead of considering all those (higher-order) mediators, we lump all possible relations, between two topics, in a single parameter, represented by their overlap in the topic space. This yields a minimal representation of the empirical data, in the lowest dimensional space.

\section{Social network's topology reflects opinion segregation}%opinion formation %opinion dynamics
%%here topology - dynamics relation. start from real world.

On social media, opinion polarization can be reflected in the topology of the corresponding social networks: The users interact more likely with peers sharing similar opinions, a situation known as \emph{echo-chambers} \cite{mcpherson2001birds,flaxman2016filter,del2016echo}.
Our model assumes that the opinion evolution is coupled to the dynamics of the underlying social network via Eqs.~\eqref{eq:vector} and \eqref{eq:homophily}. 
This mechanism yields a social network structure which is shaped by the process of opinion formation \cite{starnini2016emergence}. 
Figures~\ref{fig:networks}(a), (b), and (c) show the social networks generated by the model for the same parameters employed in Fig. \ref{fig:fig1}(a), (b), and (c), corresponding to global consensus, uncorrelated polarization, and ideological state, respectively. 
The networks result from the time-integration of the last 70 time steps of the temporal adjacency matrix $A_{ij}(t)$, once the system reaches a steady state. 
Each node corresponds to an agent $i$, size of the node is proportional to his conviction (given by $r_i$), while the color represents the opinion in the polar coordinate $\varphi_i$.

\begin{figure*}
\includegraphics[width=0.9\linewidth]{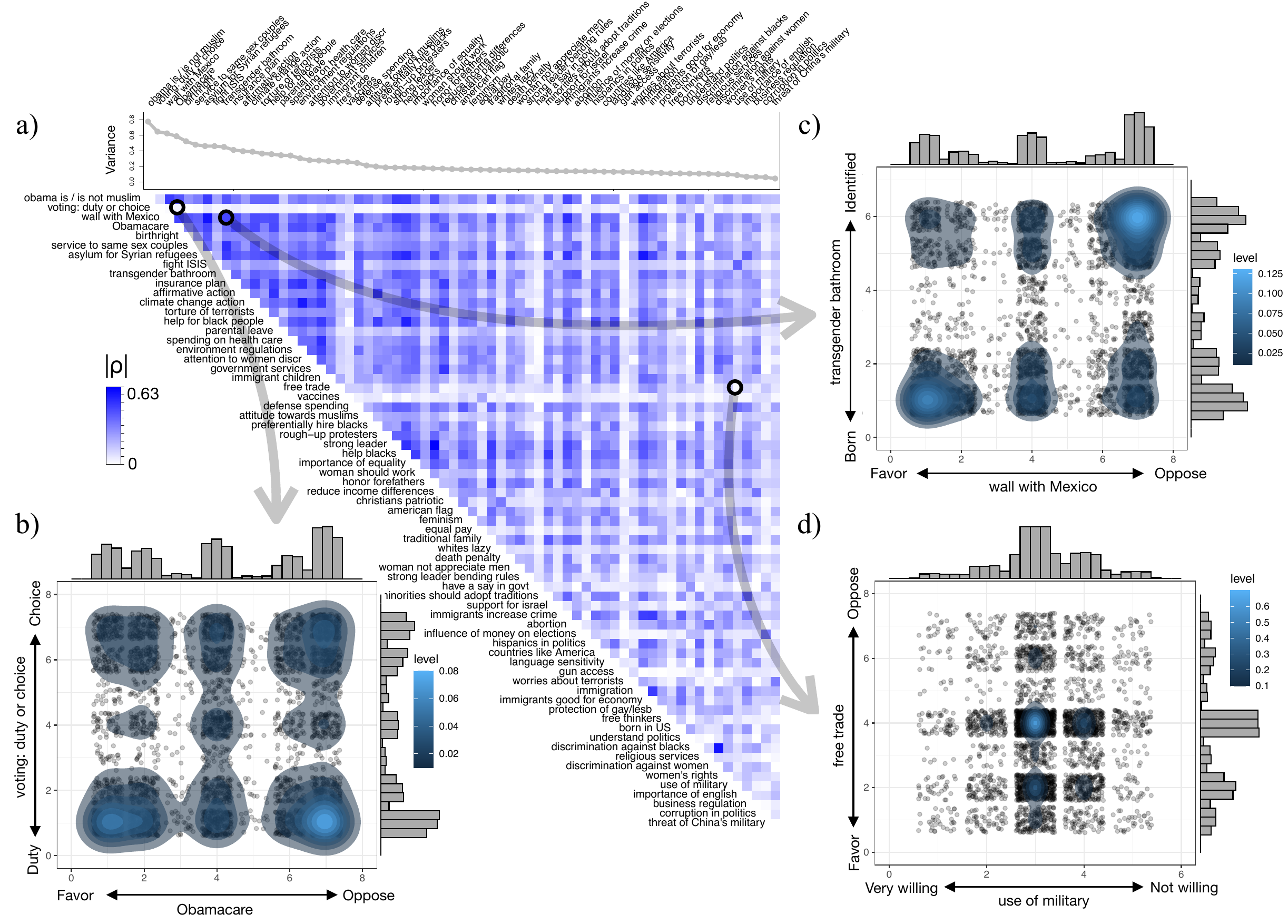}
\caption{\textbf{Responses to questions from the ANES survey.} a) Variance of all responses and absolute value of pairwise Pearson correlation.
b)-d) Scatter plots of selected pairs of questions $v$ and $z$,
where each dot represents one respondent by his/her responses to both questions.
The marginal plots represent the response distributions $P_v(x)$ and $P_z(x)$. To improve the visualization, data is jittered \cite{few2008solutions}, {uniformly distributed noise is applied, to up to 80\% of the bin-size, to avoid over-plotting of the categorical data and at the same time assure discrete separation.}
The examples are selected to represent
different combinations of response variance (opinion polarization) and response correlation: d) low variance {($\sigma^2_v=0.08$, $\sigma^2_z=0.25$)} and low correlation {($|\rho(v,z)|=0.02$)} for questions {V162176x} vs. {V161154}, denoting ANES IDs (see SM~{\cite{SM}} for a complete list of IDs);
%(``Do you favor, oppose the U.S. making free trade agreements with other countries?" vs. ``How willing should the United States be to use military force to solve international problems?");
b) high variance {($\sigma^2_v=0.58$, $\sigma^2_z=0.64$)} and low correlation {($|\rho(v,z)|=0.03$)} for {V161151x} vs. {V161114x}; and
%(``Do you consider voting a choice or duty" vs. ``Do you favor, oppose the health care reform law passed in 2010?"); 
c) high variance {($\sigma^2_v=0.62$, $\sigma^2_z=0.49$)} and high correlation {($|\rho(v,z)|=0.44$)} for  {V161228x} vs. {V161196x}.}
%(``Should transgender people have to use the bathrooms of the gender they were born as, or should they be allowed to use the bathrooms of their identified gender?" vs. ``Do you favor, oppose building a wall on the U.S. border with Mexico?"). 
\label{fig:data}
\end{figure*}

%here description of resulting networks
Fig.~\ref{fig:networks}(a) shows the system approaching global consensus. While nodes with similar opinions are more likely to be connected -- an effect caused by homophily, also in the case of low $\alpha$ -- no clear groups emerge in the  network structure. 
Fig.~\ref{fig:networks} (b) shows that in the uncorrelated polarized case, on the contrary, four groups are clearly visible, each one characterized by a different opinion (color coded as in Fig.~\ref{fig:fig1}). 
A similar situation is visible in Fig.~\ref{fig:networks}(c), depicting the ideological state, where the social network is mainly segregated into two groups, holding different opinions. 

%%new par: quanitification through comm structure
These observations can be quantified by a community detection analysis. 
Figs.~\ref{fig:networks} (d), (e), (f) show the community structure of the corresponding networks, plotted as polar bar plots, as obtained by the Louvain algorithm \cite{blondel2008fast}. 
Each community is represented as a different angle sector, which is orientated (polar angle) according to the average opinion $\langle \varphi\rangle$ within that community.
The size of the community is represented by the 
radius of each {polar} bar, while the width and color of each sector represent the average cosine similarity between nodes in that community,
%$\langle\cos(\theta)\rangle=\langle\mathbf{x}_i\cdot\mathbf{x}_j/(||\mathbf{x}_i||||\mathbf{x}_j||)\rangle$.
the mean scalar product of opinion directions calculated according to Eq. \eqref{eq:dot-product} and averaged over all pairs of agents %$(i,j)$ 
within the community.% $ \mathcal{C}$. 

%$\theta$ denotes the angle between two specific opinion vectors $\mathbf{x}_i$ and $\mathbf{x}_i$ 
%and the average is taken with respect to all pairs of agents $(i,j)$ within a community. \Note{$\leftarrow$ def. of cosine similarity}

In the global consensus case (Fig.~\ref{fig:networks}(a)), many communities are present and are rather randomly oriented. Each community is characterized by a heterogeneous spectrum of opinions, (low values of the average cosine similarity). 
On the contrary, when consensus is broken, the average opinion of the agents within each community is aligned with the dynamical attractors shown in Fig.~\ref{fig:fig1}(e) and (f).
In the uncorrelated polarized case, Fig.~\ref{fig:networks}(b), the communities are characterized by four typical average opinions, corresponding to the four colors shown in Fig.~\ref{fig:fig1}(e).
Within each community, opinions are very similar, with large values of the average cosine similarity.
In the ideological phase -- Fig.~\ref{fig:networks}(c), communities are characterized by only two typical averages opinions, cf. Fig.~\ref{fig:fig1}(f), and a strong homogeneity of opinions (very high average cosine similarity).

%The relation between the agents' behavior in the space of interactions (observed in the network topology) and the space of opinions (observed in the underlying ideological space) can be quantified by the correlation between the distances in the two spaces. When the system reaches the steady state, for each pair of agents $i$ and $j$ we compute their distance in the opinion space, $d_o(i,j) \equiv d(\mathbf{x}_i(t\rightarrow\infty),\mathbf{x}_j(t\rightarrow\infty))$ and in the network space $d_n(i,j)$. The latter distance is obtained by mapping all nodes into a low dimensional space with the embedding algorithm node2vec \cite{grover2016node2vec}, which maximizes the likelihood of preserving the neighborhoods of the nodes. Within such low-dimensional space, the network distance between two nodes corresponds to the Euclidean distance of the corresponding embedded node vectors. Figs.~\ref{fig:networks}(d), (e)  and (f) show a density scatter plot of the network $d_n$ and opinion distance $d_o$ for all pairs of agents.One can see that, while in the case of global consensus these quantities are rather independent (Figs.~\ref{fig:networks}(d)), when polarization emerges they become clearly correlated. For both cases of overlapping and non-overlapping topics, the network distance $d_n$ increases quickly as soon as the opinion distance $d_o$ grows. As $d_o$ grows larger, $d_n$ reaches a plateau. This clearly shows \ldots \MS{remark main point.} \FB{improve and explain more}.

\section{Comparison with empirical data}
The presence of three different scenarios suggested by our model can be compared with empirical data. 
In what follows, we investigate the degree of polarization and correlation between opinions with respect to different topics using data collected by the \emph{American National Election Study} (ANES).
The ANES study is a continuation of a series of surveys run since 1948, with the main objective of analyzing public opinion and voting behavior in the U.S. presidential elections by interviewing a representative sample of U.S. citizens.
The ANES data have been proven to be suitable for a variety of research purposes, ranging from examining the {driving forces} for public attitudes towards specific topics like immigration \cite{citrin1997public}, observing longitudinal developments of trust in the American government \cite{poznyak2014trust}, {to} characterizing long-term trends of polarization \cite{lelkes2016mass,baldassarri2008partisans}. 
%The data also allows for measurements of particular interest for our modelling approach: the correlations of opinions between different topics (c.f., issue constraints/alignment). Some of which were found to have modestly grown in the last decades \cite{baldassarri2008partisans}.

%%%%THIS IS FOR SM
%and classify empirically different types of two dimensional opinion states, which are categorized in terms of opinion correlations and opinion polarization. Those states render compatible with our modeling framework, allowing to draw conclusions about the relation of involved issues in topic space. \Note{More specific, what is the main finding? Which conclusions can be drawn by looking at the data, and considering the model?}

For our analysis, we select a total of $67$ questions with overall $253984$ valid responses from the 2016 ANES. See Appendix~\ref{app:emp-data} for details on the selection criteria and the SM~{\cite{SM}} for a complete list of analyzed questions.
%For our purpose, we consider the 2016 ANES, including a total of $1842$ questions to $4270$ respondents. 
Respondents are assigned an individual ID, such that their answers to different questions can be related to each other.
%to each other making possible to relate their opinions with respect to different topics.
%This is where a direct relation to our model is established: respondents express their opinions towards multiple topics under consideration. 
%The ANES data sets contain a very rich repertoire of questions (overall, $1842$ items), which are to be answered in a variety of different modes by the respondents.
%For our analysis, we exclude questions with free-text answers and only take into account multiple-choice questions, which quantify the extent of approval or disapproval of the respondent with respect to a certain issue. 
%Specifically, we considered only questions whose response scale can both distinguish between the qualitative stance towards a topic (favor or oppose) and quantifies the conviction of the respondent toward one of the sides of the issue (e.g., favor a great deal, \ldots, neutral, \ldots, strongly oppose), with at least a 4-point scale. 
%This makes the responses readily comparable across questions, since in this case, the opinions of respondents can be distributed on a (quasi) linear scale. 
%We also excluded questions regarding the political stance of the respondents towards presidential candidates or parties. 
%(not counting exclusions, drop outs, etc.), coming from $4270$ individuals.
In the following, we will focus on two key features of the ANES data: i) the distribution of responses with respect to each question, quantifying the degree of polarization or consensus toward a certain topic, and ii) the correlation between responses with respect to different {pairs of} questions, revealing which issues are aligned and thus contribute to an ideological state. 

A schematic illustration of the subset of considered issues is given in Fig.~\ref{fig:data}. 
On top of Fig.~\ref{fig:data}(a), we plot the variance $\sigma^2_u(x)$ of the response distribution to question $u$. 
Questions are sorted according to $\sigma^2_u(x)$ in descending order, from questions with most polarized responses to less polarizing ones.
While for the majority of questions (on the right side of the marginal plot) a consensus looks achievable, few questions (on the left side of the plot) are strongly polarized, such as the question of whether ``voting is a duty".
Panel (a) shows the correlation matrix of the responses, sorted according to their variance.
The cell ($u,v$) is color coded according to the absolute value of the Pearson correlation between the opinion distributions $P_u(x)$ and $P_v(x)$, $\vert \rho(u,v) \vert$.
The full distribution of correlation values for all investigated pairs of questions is reported in the SM~\cite{SM}. The average correlation value is 0.2, but the distribution is broad: some pairs of questions are weakly correlated, while others are strongly so.
Note that although there is a small dependence of the strength of correlation on the variance (slight decay of correlation towards the bottom right), both large and small correlation values can be observed in all parts of the matrix.
%\MS{Relation between variance (polarization) and  correlation.}

Panels (b)-(d) of Fig.~\ref{fig:data} show three prototypical cases corresponding to the three steady states found in our model: consensus (d), polarization (b) and ideological state (c). 
The first case corresponds to questions whose responses are both peaked around a neutral opinion, with a low variance of the opinion distribution. This case is shown in Fig.~\ref{fig:data} (d) by questions ``Do you favor, oppose the U.S. making free trade agreements with other countries?" {(answer on a 7 point scale)} vs. ``How willing should the United States be to use military force to solve international problems?" {(5 point scale)}.
Fig.~\ref{fig:data} (b) shows the questions 
``Do you consider voting a choice or duty" {(7 point scale)} vs. ``Do you favor, oppose the health care reform law passed in 2010?" {(7 point scale)} (obamacare law), which have polarized responses that are not correlated.
Finally, the case of polarized opinions that are strongly correlated is shown in Fig.~\ref{fig:data} (c), with the questions ``Should transgender people have to use the bathrooms of the gender they were born as, or should they be allowed to use the bathrooms of their identified gender?" {(6 point scale)} vs. ``Do you favor, oppose building a wall on the U.S. border with Mexico?" {(7 point scale)}.

One may expect strong opinion correlations only for a pair of questions dealing with very similar topics, such as the one stated in our initial example, about transgender bathrooms and same-sex marriage, which seem intimately related to each other. In the SM~{\cite{SM}} we show that the responses to these questions are indeed strongly correlated.
The question about building the wall to Mexico, however, seems to be rather unrelated to the issue of transgender bathrooms,
%, such as is related to questions about Syrian refugees, as they both refer to immigration.
%we find a high correlation questions are indeed correlated.
so that the high correlation in  Fig.~\ref{fig:data}(c) comes as a surprise. 
This is not a rare example, and three more are shown in Fig.~S3(c)-(f) of the SM~\cite{SM}.

{While indeed the correlation between opinions with respect to similar topics may seem trivial, the emergence of such correlation in the case of unrelated topics is more puzzling.
There might be several confounding factors responsible for this correlation.
%, such as the issues of the wall with Mexico and transgender bathrooms. 
Our model provides a twofold framework to approach this. 
On the one hand, one might consider a low-dimensional representation in which all possible relations between these two topics are encoded into a single parameter, represented by their overlap in the topic space.
%we may find a minimal description and lump all possible relations between two topics into a single parameter, represented by their overlap in the topic space.
As suggested by our mean-field analysis, opinion correlation can emerge easier when topics are more controversial, due to social re-enforcement. 
This is shown in Fig.~\ref{fig:stability}: the phase transition between consensus and ideology is critically determined by the controversialness parameter $\alpha$, as also indicated by Eq. \eqref{eq:stability}. 
Two pairs of topics with similar overlap $\cos(\delta)$ but different controversialness $\alpha$ are predicted to be in different phases: consensus (low $\alpha$) vs ideology (large $\alpha$). Hence, the emergence of correlations between opinions with respect to topics with small overlap is driven by social reinforcement in the case of large controversialness.
On the other hand, 
%Beyond this minimal two-dimensional representation of empirical data in the lowest dimensional space possible, 
%Beyond, the On the other hand, as discussed in Sec.~\ref{sec:higher_dimensions}, 
one might also explicitly consider
%want to explicitly consider 
a higher dimensional space, in which additional topics, not observed in the empirical data, are included.
%that are potentially not observed in empirical data, such as the ANES survey. 
This may give rise to polarized ideological states also for independent, i.e. orthogonal, topics, as we have shown in Sec.~\ref{sec:higher_dimensions} for three dimensions.}
%\tr{NEW+++++++++++}

%Our model offers an socio-mechanistic explanation for such high correlations of remotely overlapping topics: even if the overlap is small, the social interactions collectively re-enforce the correlation of overlapping topics above a critical angle $\delta$ (see Fig.~\ref{fig:stability}).

%This goes to discussion
%All these observations can be covered with our relatively simple model, assuming only a few general mechanisms, social influence and non-orthogonal topics. Thus, the interpretation we offer, relies solely on the inherent thematic overlap of topics, without assuming differences in the agents or in their opinion formation process.

% Nice formulation!
%Our models are obviously highly stylized. But their main elements, small-world network structures and social influence and selection processes, capture important features of intercultural contact in the social world.

\section{Conclusions}
To sum up, we proposed a simple model able to reproduce crucial features of opinion dynamics as measured in survey data, such as consensus, opinion polarization, and correlation of opinions on different issues, i.e. ideological states. 
Our model is based on three main ingredients, inspired by empirical
evidence:  i) The opinion formation is driven by time-varying, homophilic social interactions among the agents, ii) agents sharing similar opinions can mutually reinforce each other's stance, and iii) opinions lay in a multidimensional space, where topics form a non-orthogonal basis (i.e. they can overlap) and can be controversial.
{Opinion correlations emerge as soon as the assumption of an orthogonal basis is relaxed and topics are allowed to partly overlap. Ideological states appear as a purely collective phenomenon without explicit assumptions of individual attributes of agents favoring one partisanship over another}.
We analytically and numerically characterize the transitions between the three states, consensus, polarization, and ideology, in dependence on the controversialness and overlap of the topics discussed. 
The model describes the possibility of strong correlations between opinions with respect to rather unrelated topics provided they are controversial enough, which prediction is corroborated by empirical data of questionnaire surveys.

%\rev{The topic overlap introduced here is not a purely theoretical concept with a geometrical interpretation. On the contrary, it would be interesting to devote further research to close the gap between two independent empirical observations: i) the correlation between opinions with respect to different topics (quantified by surveys or extracted from online social media), and ii) the thematic overlap between these two topics, that could be determined by, for instance, methods from natural language processing such as word embedding of text documents. (IN THE REBUTTAL, SAY WE ADDED THIS THANKS TO THE DISCUSSION WITH THE REFEREES BLABLA)}

%%%Pitch
%Empirical data shows that also opinions with respect to rather unrelated topics can be strongly correlated. 
%This finding poses a challenge from a modelling perspective: how such correlation can emerge, without assuming it a priori in the behavior or individual preferences of the agents or in a preexisting social structure? 
%Our model proposes a reinforcement mechanism driven by homphilic social interactions that leads to the formation of ideological states, even between topics with small overlap but that are sufficiently controversial.
%We solidify these intuitions in a formalism where opinions {evolve in a multidimensional space, spanned by non-orthogonal topics}.
%This approach was inspired by  skew coordinate systems that have  been  recently  proposed  to solve some well-known issues of classical vector space models for representing text document.

Of course, our work comes with limitations. 
With respect to the modeling perspective, it is important to note that our model is based on a minimal number of assumptions. 
It disregards some empirical features of social interactions such as individual preferences of the agents. 
This is, however, a necessary trade-off between including realistic features of human behavior and the need to keep the model as simple as possible and the number of parameters small. 
%{A data set which comprehends of a large set of topics, such as the ANES, and includes both temporal opinion and network information is absent, to the best of our knowledge, and would be quite difficult to collect, not least due to privacy constraints. Thus the empirical validation of the role of social interactions and their temporal component (evolution of opinions) are not possible at this time.} 
{With respect to the empirical validation, the direct tests about the role of social interactions and the impact of the temporal dimension (evolution of opinions) are not possible on the ANES data set. 
Indeed, a data set which is comprehensive of a large set of topics, such as the ANES, and includes the aforementioned temporal and network information is absent, to the best of our knowledge, and would be quite difficult to collect, also for privacy constraints.} 
The ideal venue to build such data sets could be online social media, where users can take advantage of anonymity in expressing their opinions and social interactions could be reconstructed.   
We left the design of such as study as important future work. 
{The proposed framework also suggests another interesting direction for future work: to investigate the relation between opinion polarization and issue alignment, whose empirical evidence remains unclear \cite{baldassarri2008partisans}.}
%Finally, it would be extremely interesting to directly quantify topic overlaps in surveys, such as the ANES. This challenge could be addressed by topic modeling of large data sets related to the topics under consideration, such as news articles, and then projecting the trained model (i.e., the topics forming the basis of the space) to the survey data under consideration. 

{Finally, the topic overlap introduced here is not a purely theoretical concept with a geometrical interpretation.
On the contrary, it would be interesting to devote further research to close the gap between two independent empirical observations: i) the correlation between opinions with respect to different topics (quantified by surveys or extracted from online social media), and ii) the thematic overlap between these two topics. This latter challenge could be addressed by topic modeling of large data sets related to the topics under consideration, such as news articles, and then projecting the trained model (i.e., the topics forming the basis of the space) to the survey data under consideration.}

\section*{Acknowledgements}
This work was developed within the scope of the IRTG
1740/TRP 2015/50122-0 and funded by the DFG and
FAPESP.

\appendix
\section{Numerical simulations}\label{app:num-sim}
For the numerical simulations of Eqs.~\eqref{fig:fig1} we set the basic simulation parameters to the following values: $N=1000$, $T=2$, $\beta=3$, $K=3$. The parameters of the basic AD model are set to ($m=10$, $\epsilon=0.01$, $\gamma=2.1$), and the activities of agents, $a_i$, are drawn from the distribution $F(a) = \frac{1-\gamma}{1-\epsilon^{1-\gamma}}\,a^{-\gamma}$. The results depicted in Figs.~\ref{fig:fig1}-\ref{fig:networks} differ with respect to the values of $\alpha$ and $\delta$, as reported in the captions and the main text. 
The opinions are initialized as two and three dimensional Gaussian distributions with mean and variance of $\mu=0$ and $\sigma^2=2.5$, respectively, and there are no connections between agents.

The temporal network $A_{ij}(t)$ and the opinion vectors $\mathbf{x}_i$ are updated at each time step $t$ as follows.
\begin{itemize}
    \item[1)]  {Initially, {in each time step,} the system consists of $N$ disconnected nodes, and hence} the temporal adjacency matrix $A_{ij}(t)$ is the zero matrix. {Subsequently,} each agent $i$ is activated with probability $a_i$\,.
    \item[2)]  Each active agent $i$ contacts $m$ distinct agents, where the probability that agent $i$ contacts agent $j$ is given by $p_{ij}$, cf. Eq.~\eqref{eq:homophily}. The opinion distance $d(\mathbf{x}_i, \mathbf{x}_j)$, between agents $i$ and $j$, is computed involving Eq.~\eqref{eq:dot-product}. {Note that agent $i$ samples $m$ links based on $p_{ij}$ without replacement, such that agent $i$ can contact agent $j$ only once per time step.} The elements of the temporal adjacency matrix $A_{ij}(t)$ are set to $A_{ij}(t) = A_{ji}(t)=1$ if agent $i$ contacts agent $j$, or vice-versa. 
    \item[3)] After the temporal adjacency matrix $A_{ij}(t)$ is generated, for each agent $i$ the aggregated social input coming from its neighbors is computed and the opinion vector $\mathbf{x}_i(t+dt)$ is updated by numerically integrating Eq.~\eqref{eq:vector} using an explicit Runge-Kutta 4th order method \cite{press1989numerical} with $dt=0.01$. {After the opinion vector is updated the process starts anew from 1).}
\end{itemize}

\section{Mean-field approximation}\label{app:mf-approxiation}
For an arbitrary number of topics $T$,
in case of a large number of agents ($N\gg1$) and strong homophily ($\beta\gg1$), an agent's opinions will be close to the opinions of its interaction partners, i.e. we have $x_i^{(u)}\approx x_j^{(u)}\equiv x^{(u)}$ in Eq.~\eqref{eq:vector}. 
In this approximation, the dynamics of a single agent is then effectively described solely by interactions with neighbors holding the same opinion, i.e., a self-interacting agent. 
For fast switching interactions, the average number of interactions received by an agent at each time step can approximated by ${2}m\langle a\rangle${, which is a sum of two contributions. First, the average number of links an agent generates upon activation is $\langle a\rangle m$, and a second contribution, which stems from links expected to be received by agent $i$ from all other agents, $\langle a\rangle m=\langle a\rangle\sum_{j=1}^N\frac{m}{N}$}. Hence, Eqs.~(1) reduce to 
\begin{align}\label{eq:single_agents_all_topic}
\dot{x}^{(v)} &= - x^{(v)} + {2}K m\langle a\rangle \tanh\left(\alpha [\mathbf{\Phi x}]^{(v)}\right)\,,
\end{align}
which describes the opinion dynamics of agents, depending on the topic overlap matrix $\mathbf{\Phi}$.

The relation between the controversialness $\alpha$ and the topic overlap $\cos(\delta)$, marking the transition between a global consensus and the emergence of opinion polarization, can be derived using the Jacobian of Eq.~\eqref{eq:single_agents_all_topic}. To capture the transition analytically, we additionally assume that all pairwise topic overlaps are equal, i.e. the angles between topics are $\delta_{uv}=\delta\quad\forall u,v$. 
The Jacobian of Eqs.~\eqref{eq:single_agents_all_topic} evaluated at $\mathbf{x}=0$ yields
\begin{equation}
 \mathbb{J}(\mathbf{0})=
\begin{pmatrix}
    -1+\Lambda\alpha & \Lambda\alpha\cos(\delta)&\dots& \Lambda\alpha\cos(\delta)\\
   \Lambda\alpha\cos(\delta)& -1+\Lambda\alpha&\ldots& \Lambda\alpha\cos(\delta)\\
    \vdots &\vdots &\vdots&\vdots\\
    \Lambda\alpha\cos(\delta)  &\Lambda\alpha\cos(\delta)&\ldots& -1+\Lambda\alpha\\
\end{pmatrix},
\end{equation}
where we have defined $\Lambda ={2}Km\langle a\rangle$ for brevity. The largest eigenvalue of $\mathbb{J}(\mathbf{0})$, $\lambda_\mathrm{max}$, is given as 
\begin{align}\label{eq:largest-eigval}
    \lambda_\mathrm{max} = (T-1)(-1+\Lambda\alpha)+\Lambda\alpha\cos(\delta).
\end{align}
If $\lambda_\mathrm{max}<0$ the full consensus is stable. Finally, setting Eq.~\eqref{eq:largest-eigval} to zero and solving for $\alpha$ yields
\begin{align}\label{eq:stability-general}
    \alpha_c=\frac{T-1}{{2}Km\langle a\rangle [T-1+\cos(\delta)]}\,,
\end{align}
which relates the critical controversialness $\alpha_c$ to the topic overlap $\cos(\delta)$ for an arbitrary number of topics $T$. 

For the sake of simplicity, in the paper we mainly consider the case of two topics. 
Setting $T=2$ in Eq.~\eqref{eq:stability-general} yields Eq.~\eqref{eq:stability}.
In this case,  Eqs.~\eqref{eq:single_agents_all_topic} is reduced to  the following non-linear system of equations 
\begin{align}
\dot{x}^{(u)} &= -  x^{(u)} +  2Km\langle a\rangle\tanh\left(\alpha \left[ x^{(u)}+\cos(\delta) x^{(v)}\right]\right)\nonumber  \\
\dot{x}^{(v)} &= - x^{(v)} + 2Km\langle a\rangle\tanh\left(\alpha \left[\cos(\delta)x^{(u)}+x^{(v)}\right]\right)\,,\label{eq:mean-field-two-topics}
\end{align}
which give rise, for $2Km\langle a\rangle=1$, to the attractor dynamics depicted in subpanels (d)-(f) of Fig.~\ref{fig:fig1}.
%To compare the dynamics of the mean-field approximation with the stochastic dynamics depicted in Fig.~1(a)-(c) we consider the case two topics. 

The stability regions in the $\cos(\delta)$-$\alpha$ space, depicted in Fig.~\ref{fig:stability}, are computed based on the Jacobian of Eqs.~\eqref{eq:mean-field-two-topics}. 
While the critical controversialness (black dashed line in Fig.~\ref{fig:stability}) is analytically given by Eq.~\eqref{eq:stability}, the regions of stability for correlated and uncorrelated polarization must be determined numerically. 
In the mean-field approximation, we define as uncorrelated polarized states all situations in which the system has two stable fixed points $\mathbf{x}^*$ with $[\sgn(x^{(u)*}), \sgn(x^{(v)*})]=(-,+)$ and $[\sgn(x^{(u)*}), \sgn(x^{(v)*})]=(+, -)$, respectively. The stability of these fixed points is determined numerically in a two-step procedure. 
Upon discretizing the $\cos(\delta)$-$\alpha$ plane, we first compute, for each $\{\alpha,\cos(\delta)\}$ parameter combination, the values of the two fixed points by using the Newton-Raphson method \cite{press1989numerical}. 
In a second step, we numerically determine the stability of these fixed points $\mathbb{\mathbf{x}^*}$ by computing the largest eigenvalue of  $\mathbb{J}(\mathbf{x}^*)$. 
If negative, the corresponding fixed points are stable, and the system is in an uncorrelated polarized state. Otherwise, they are unstable and the system will fall to a polarized ideological state.
%As such fixed points loose their stability, for increased values of topic overlaps, we define the state as ideology state.

{The ideological phase, depicted in the phase-space diagram in Fig.~\ref{fig:stability}, extends, for $\alpha=1$, until the line of vanishing overlaps ($\cos(\delta)=0$). At the corresponding triple point, at $\cos(\delta)=0$ and $\alpha=1$, infinitely small topic overlaps will give rise to ideological states. This can be  understood examining the non-trivial fixed-point solutions to  Eqs.~\eqref{eq:mean-field-two-topics} for $2Km\langle a\rangle=1$ and $\alpha=1$. Setting Eqs.~\eqref{eq:mean-field-two-topics} to zero, taking $\tanh^{-1}(\dots)$ of both sides, and Taylor expanding the non-linearity up to third order, yields
\begin{align}
 \frac{\left(x^{(u)}\right)^3}{3}  &\simeq    \cos(\delta) x^{(v)} \nonumber   \\
 \frac{\left(x^{(v)}\right)^3}{3} &\simeq   \cos(\delta)x^{(u)}\,. \label{eq:taylor_triple_point}
\end{align}
%Taylor expanding the right-hand sides of the latter equations up to third order, gives $\frac{\left(x^{(1)}\right)^3}{3} =   \cos(\delta) x^{(2)}$ $\frac{\left(x^{(2)}\right)^3}{3} =   \cos(\delta)x^{(1)}\label{eq:ideol2}$
The latter relations suggest that, for $\cos(\delta)>0$, non-vanishing solutions (${x}^{(u)*},{x}^{(v)*}\ll 1$) yield equal opinion stances, with respect to both topics, i.e. we have $[\sgn(x^{(u)*}), \sgn(x^{(v)*})] = (+,+)$ or $(-,-)$. In particular, the solutions behave as $(x^{(u)*},x^{(v)*})=\sqrt{3\cos(\delta)}(1,1)$, and $(x^{(u)*},x^{(v)*})=\sqrt{3\cos(\delta)}(-1,-1)$ close to the triple point.}

Note that for $\cos(\delta)<0$ ($\delta\in\,\,]\pi/2,\pi[$) the stability of the system is reversed giving rise to negatively correlated opinions, as shown in the SM~\cite{SM}.
However, this does not lead to qualitatively new dynamical features, {and leads to $[\sgn(x^{(u)*}), \sgn(x^{(v)*})] = (-,+)$ or $(+,-)$ close to the triple point.}
With respect to our empirical data analysis, this merely corresponds to re-formulating one of the two questions with a reversed scale. Therefore, we omit this range of negative topic overlap and focus on $\delta\in]0,\pi/2]$, i.e. positive overlaps.

\section{Empirical Data}\label{app:emp-data}
The data set analyzed for this work is the 2016 American National Election Survey (ANES) \cite{anes}. It includes a total set of 1842 questions. Each of the 4270 respondents is assigned an individual ID, which allows us to correlate responses given by a respondent to different questions. %Note however, that the questions differ with respect to their specific response type.
In order to quantify the degree of polarization and issue alignment we compute the variances of responses to single questions and the Pearson correlation coefficients $\rho$ between the responses to pairs of questions.
In the caption of Fig. \ref{fig:data} we report these values for the three examples discussed in the main text, other values can be found in the SM~\cite{SM}. 

This procedure requires a numerical scale for the responses. Therefore, we first exclude all questions with free-text answers, such as ``What kind of work did you do on your last regular job?". 
The remaining questions are multiple-choice questions, not all well suited for our purpose. We only select those questions which allow us to extract the extent of approval or disapproval of the respondent with respect to a certain issue. 
In particular, we choose questions whose response scale allows us to quantify both the qualitative stance (favor or oppose) and the conviction (e.g., favor a great deal, \ldots, neutral, \ldots, strongly oppose) of the respondent towards the issue, with at least a 4-point scale. Questions whose response scales do not ensure this or questions which do not ask about a specific opinion, such as ``Which of the following radio programs do you listen to regularly?" are excluded. In the last step, we exclude questions regarding political parties or presidential candidates. 
%Due to their direct link to politics, we consider most of such questions as mutually exclusive (extremely large topic overlap), and expect trivial correlations in such cases.
These selection criteria reduce the 2016 ANES data set to a total of 67 questions, depicted in Fig.~\ref{fig:data}. We report the complete list of selected questions in the SM~\cite{SM}, together with the question IDs to locate them in the data set provided by Ref.~\cite{anes}.

\widetext
\newpage
\begin{LARGE}
\begin{center}
  \textbf{Supplementary Material } 
\end{center}
\end{LARGE}

\begin{center}
{\bf Different controversialness of topics and negative topic overlaps}
\end{center}

\begin{figure}[h!]
\includegraphics[width=\textwidth]{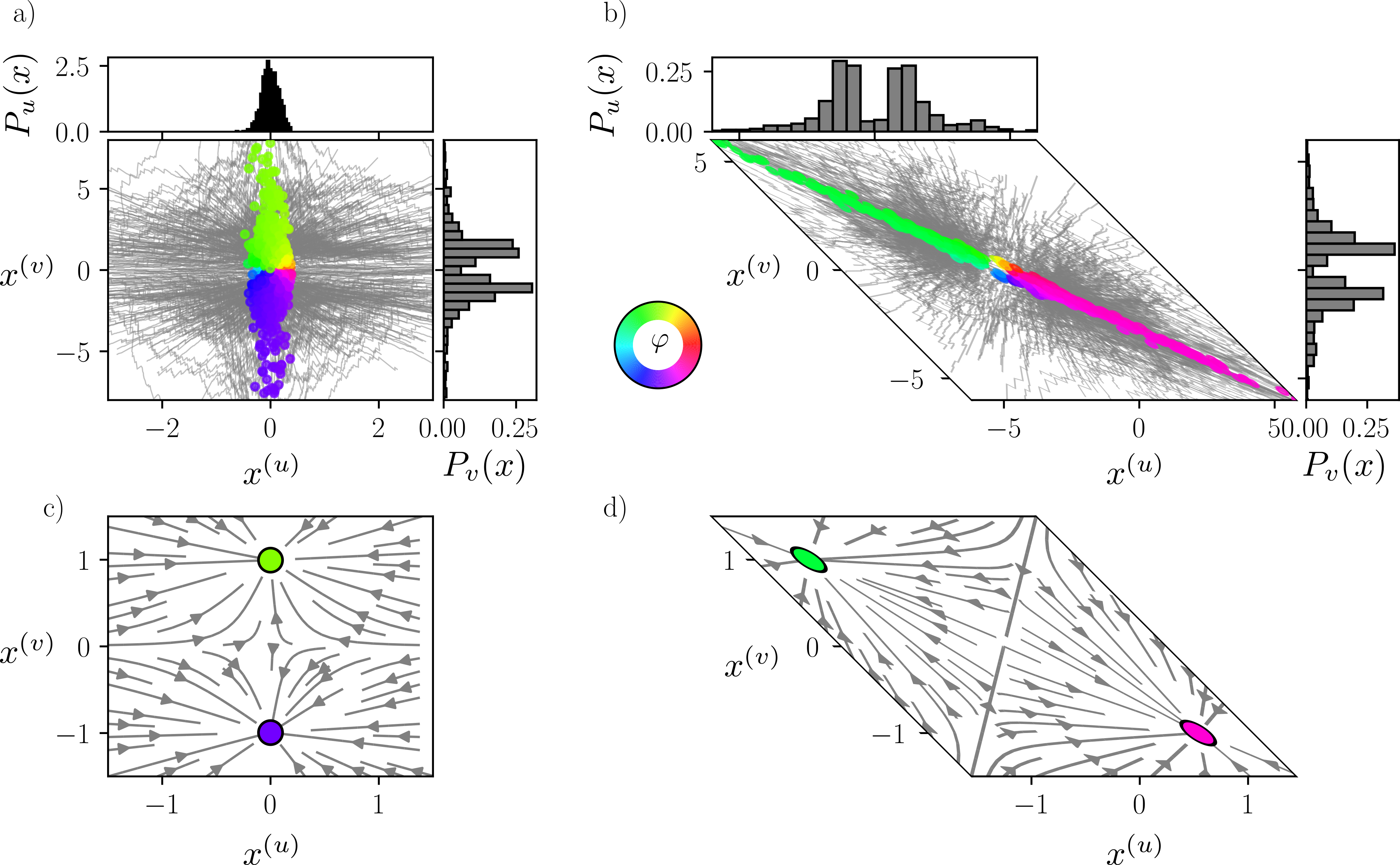}
\caption{Simulation results of the full model and the corresponding mean-field approximations for $\alpha_u=0.05$, $\alpha_v=3$ in panels  (a) and (c), $\delta=\pi/2$ and $\alpha_u=\alpha_v=3$, $\delta=3\pi/4$ in panels (b) and (d). 
All remaining parameters were set as in Fig.~1 of the main text: $N=1000$, $K=3$, $\beta=3$.}
\label{fig:different-alphas-deltas}
\end{figure}

Topics can be characterized by different values of controversialness. To account for this we generalize Eqs.~(3) of the main text to  
\begin{align}
\dot{x}_{i}^{(u)} &= - x_{i}^{(u)} + K\sum_j A_{ij}(t) \tanh\left[\alpha_u (x_{j}^{(u)}+\cos(\delta)x_{j}^{(v)})\right]
%\label{eq:T2_x} 
\nonumber \\
\dot{x}_{i}^{(v)} &= - x_{i}^{(v)} + K\sum_j A_{ij}(t) \tanh\left[\alpha_v (\cos(\delta)x_{j}^{(u)}+x_{j}^{(v)})\right]%\label{eq:T2_y}
\,,
\label{eq:different-alphas}
\end{align}
where $\alpha_u$ and $\alpha_v$ denote the controversialness values of topics $u$ and $v$, respectively. Here we briefly discuss a case of different $\alpha$ values ($\alpha_u\neq\alpha_v$), where the dynamics with respect to each topic strongly depends on the respective controversialness, cf. Fig.~\ref{fig:different-alphas-deltas}(a). Due to the small value of $\alpha_u(=0.05)$, agents {reach a} consensus on topic $u$, while polarization emerges {with respect to topic} $v$, as $\alpha_v(=3)$ is large. A similar behavior arises for the mean-field approximation, with two stable fixed points at $\mathbf{x}^*\simeq (0,\pm 1)$, cf. Fig.~\ref{fig:different-alphas-deltas}(c). Similar states can be found in the ANES data set, e.g. see Fig.~\ref{fig:other_ex}(a), where the responses with respect to one issue ("attitude towards Muslims") show a (neutral) consensus-like situation, while answers with respect to the second question ("service to same-sex couples") are strongly polarized.

As we discuss in the Methods section, the dynamics towards polarized ideological states is reversed for negative overlaps, $\cos(\delta)<0$, i.e. for topic angles $\delta\in\,\,]\pi/2,\pi[$. In Fig.~\ref{fig:different-alphas-deltas}(b) such a situation is depicted for $\delta=3\pi/4$, which corresponds to the mirrored state emerging for $\delta=\pi/4$, where opinions show strong negative correlations $\rho(x^{(u)},x^{(v)}) \simeq -1$. This behavior is also reflected in the attractor dynamics of the mean-field approximation, depicted in Fig.~\ref{fig:different-alphas-deltas}(b).

\newpage 
\begin{center}
{\bf Dynamics for low homophily}
\end{center}
{For small $\beta$ one-sided radicalized states emerge in the case of sufficient $K$ and $\alpha$ as observed for the one-dimensional model introduced in Ref.~\cite{PhysRevLett.124.048301}.} 

{
Accordingly, the multidimensional model shows different dynamics depending on the value of $\beta$, which we explored on the whole range $\beta \in [0,3]$. Here we complement the analysis for high homophily, in simulations and the mean-field approach (main text), by three additional cases of smaller values of $\beta<3$. While for vanishing homophily ($\beta=0$) the system directly enters a one-sided radicalized state, increasing $\beta$ changes this picture.  For $\beta=1$ one observes that the system also approaches a one-sided radicalized state, however, the process is slowed down by reinforcing interactions among like minded-agents. For $\beta=1.25$ the system indeed shows a polarized state, which is, however, different from the one emerging for very high homophily, i.e. $\beta=3$. The marginal opinion distributions do not have a pronounced bi-modal shape, which is also reflected on the network level:  groups holding the same opinion (depicted with the same color) are less pronounced as communities with respect to the $\beta=3$ case. Furthermore, for low homophily the trajectories of opinions (grey lines) fluctuate stronger than for $\beta=3$. This is due to social interactions being more heterogeneous, in the sense that agents are influenced also from opinions different from their owns.}

%For small $\beta$ one-sided radicalized states emerge in the case of sufficient $K$ and $\alpha$ -- as observed for the one-dimensional model introduced in Ref.~\cite{PhysRevLett.124.048301}. Without (or low) homophily, agents choose their interaction partners randomly, i.e. without a preference for agents having similar opinions. Accordingly, the opinion re-inforcement mechanism, which is necessary to stabilize polarized opinion states is inhibited, and all agents are rapidly absorbed by an opinion state, characterized by identical stance combinations for all agents. In the two dimensional case, discussed in the main text, those stance combinations are: $(-,+)$, $(-,-)$, $(+,-)$ and $(+,+)$, which was realized -- by chance -- in the simulation depicted in Fig.~\ref{fig:no-homophily}(b). In Fig.~\ref{fig:no-homophily}(a) we show for comparison the case discussed in the main text for high homophily ($\beta=3$). These different behaviors are reflected in the corresponding network structures shown in the bottom panels of Fig.~3, with four (panel (a)) and only a single opinion cluster (panel (b)), respectively.}

\begin{figure}[h!]
\includegraphics[width=\textwidth]{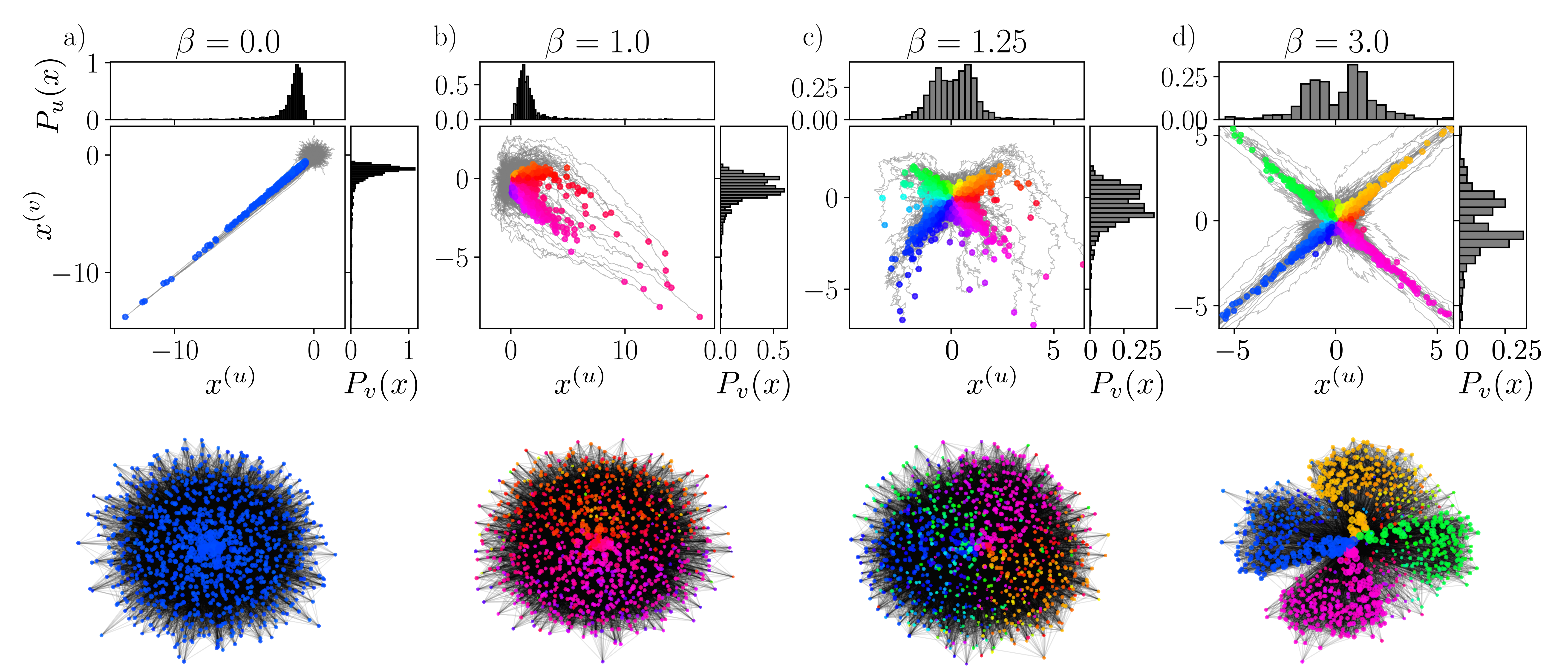}
\caption{{Dynamics for low homophily. Panels (a)-(d) show opinion evolution and the aggregated networks obtained for increasing values of $\beta=[0,1,1.25,3]$, obtained from numerical simulations of Eqs.~(4) of the main text. The remaining model parameters are set as in the main text, i.e. $N=1000$, $m=10$, $K=3$, $\alpha=3$, $\varepsilon=0.01$ and $\gamma=2.1$\,.}}
\label{fig:no-homophily}
\end{figure}

\begin{figure}[h!]
\includegraphics[width=\textwidth]{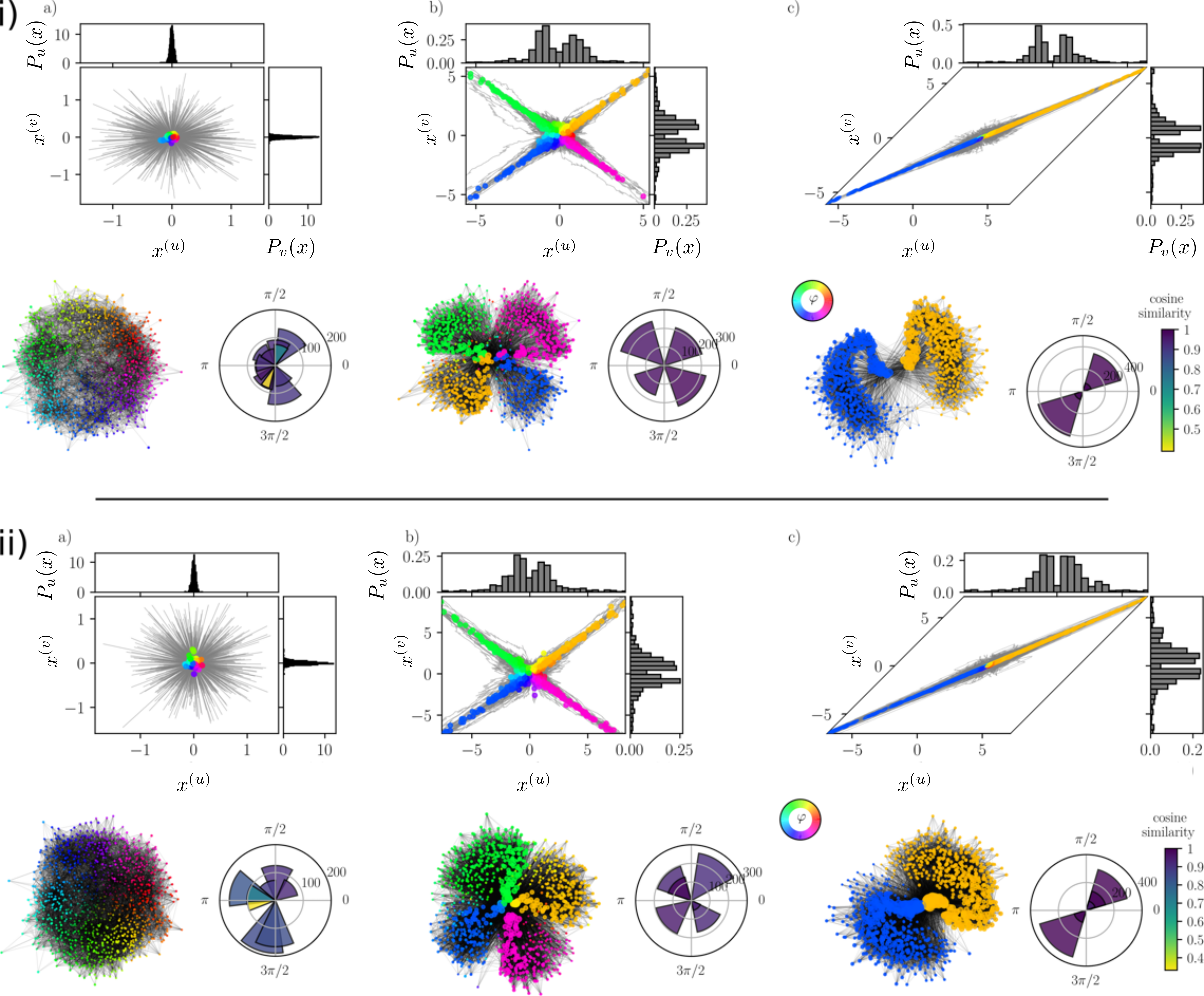}
\caption{Parameters of the activity-driven model: i) $\epsilon=0.01$, $\gamma=2.5$, $m=10$, and ii) $\epsilon=0.005$, $\gamma=2.1$, $m=20$. The values of the controversialness, homophily and topic overlap are chosen as in Figs.~1 and 4 of the main text, i.e.: $\alpha=0.05$, $\beta=3$, $\cos(\delta)=\pi/2$ (panel (a)), $\alpha=3$, $\beta=3$, $\cos(\delta)=\pi/2$ (panel (b)) and $\alpha=3$, $\beta=3$, $\cos(\delta)=\pi/4$ (panel (c)).}
\label{fig:AD_varied}
\end{figure}

\newpage
\begin{center}
{\bf Robustness of the model}
\end{center}
In this section we confirm the robustness of the model with respect to the parameters of the activity-driven {model}. While the values for homophily, controversialness and topic overlap are chosen as in the main text, below we depict simulation results for additional values of $\epsilon$, $\gamma$ and $m$. In the spirit of Fig.~1 and Fig.~4 of the main text, we show in  Fig.~\ref{fig:AD_varied} results for i) $\epsilon=0.01$, $\gamma=2.5$, $m=10$, and ii) $\epsilon=0.005$, $\gamma=2.1$, $m=20$\,. Importantly, all crucial aspects of the model are retained. For small values of $\alpha$ a global consensus emerges, cf. panels (a). For larger values of the controversialness ($\alpha=3$) the opinion distribution polarizes in an uncorrlated way, in panels (b) and as a strongly correlated ideological state in panels (c). For both additional parameter combinations, this behavior is also reflected in the network structure. 

\newpage
\begin{center}
{\bf Additional information on ANES data}
\end{center}
\begin{figure}[h]
\includegraphics[width=.7\linewidth]{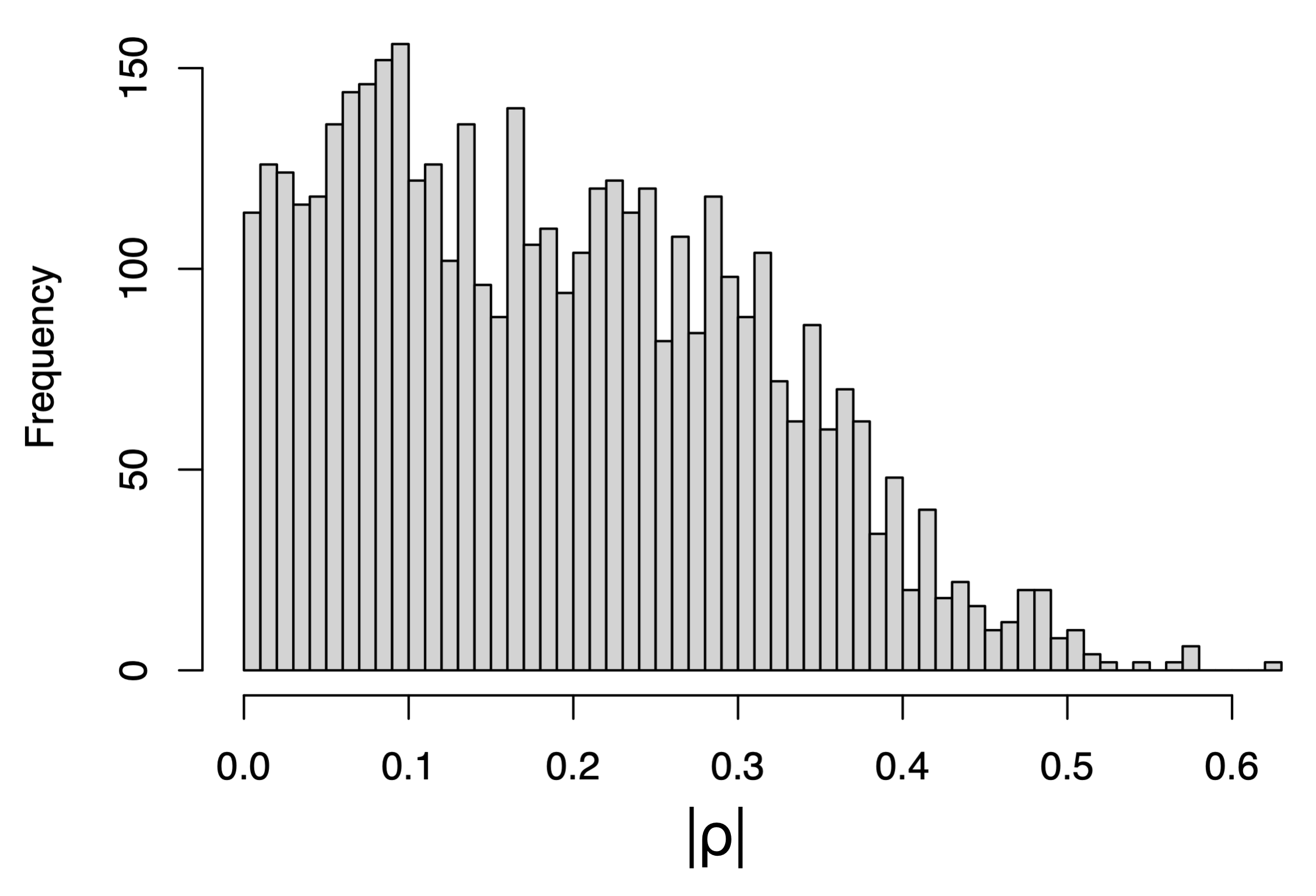}
\caption{Distribution of the Pearson correlation values between all 67 selected questions.}
\label{fig:corr_distr}
\end{figure}

\begin{table}[h]
\begin{tabular}{||c | c | c | c | c ||}
 \hline
 Question tuple $(u,v)$ & $\sigma_u^2$ & $\sigma_v^2$  & $|\rho(u,v)|$ & $p$--value ($\rho$) \\ [0.5ex]
\hline\hline
"Obamacare", "voting: duty or choice" & 0.5869515 & 0.6468563 & 0.03278208 & 0.03263277\\ 
\hline
"use of military", "free trade" &  0.08553101 & 0.2579248 & 0.0203656 & 0.2235222\\ 
\hline
"wall with Mexico", "transgender bathroom" & 0.6236126 & 0.4494401&0.4497568&0.0\\
\hline
"attitude towards muslims", "services to same-sex couples" & 0.2102911  & 0.4781575 & 0.2022116 & 0.0 \\
\hline
"services to same-sex couples", "transgender bathroom"  & 0.4781575 & 0.4494401 &0.5041266&0.0\\
\hline
"environment regulations", "insurance plan" & 0.3026623 & 0.4105342 &  0.5030662 & 0.0\\
\hline
"climate change action", "transgender bathroom" & 0.387588 & 0.4494401 & 0.3925486 & 0.0\\
\hline
"asylum for Syrian refugees", "transgender bathroom" & 0.460483 & 0.4494401 &0.4842694&0.0 \\
\hline
"asylum for Syrian refugees", "blacks should help themselves" & 0.460483 & 0.3558101 & 0.4841681 &0.0\\
\hline\hline
\end{tabular}
\caption{\label{table:pairs-of-questions} Variances of responses to single questions  ($\sigma^2_u$, $\sigma^2_v$), and Pearson correlations between responses to both questions, $|\rho(u, v)|$, for all shown question combinations in Fig.~4 (main text) and Fig.~\ref{fig:other_ex}.}
\end{table}
\FloatBarrier

\begin{figure}
\includegraphics[width=1.\linewidth]{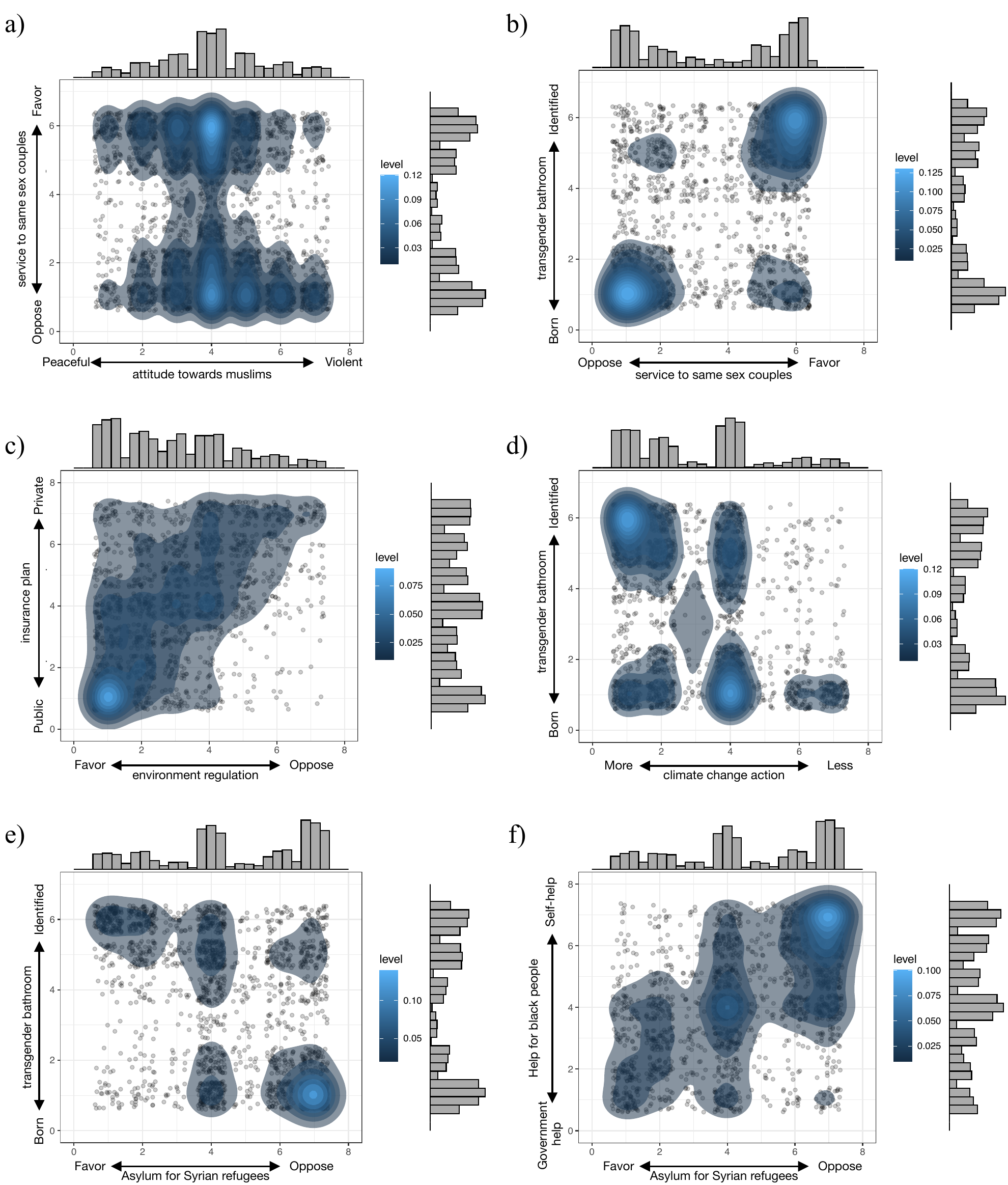}
\caption{ 
Scatter plots of selected pairs of questions, additional to those shown in the main text. 
a) Consensus (``attitude toward Muslims") vs polarization (``"service to same-sex couples"),
b) - f) polarized correlated responses with high variance and high correlation. 
Note that some topics have large overlap, like panel b), while other topics are rather unrelated, see panels c) - f).}
\label{fig:other_ex}
\end{figure}

\FloatBarrier
\begin{table}
\caption{Overview of all 67 analyzed questions, their abbreviated labels and ANES IDs.} 
\label{tab:my table} 
\includegraphics[width=\textwidth]{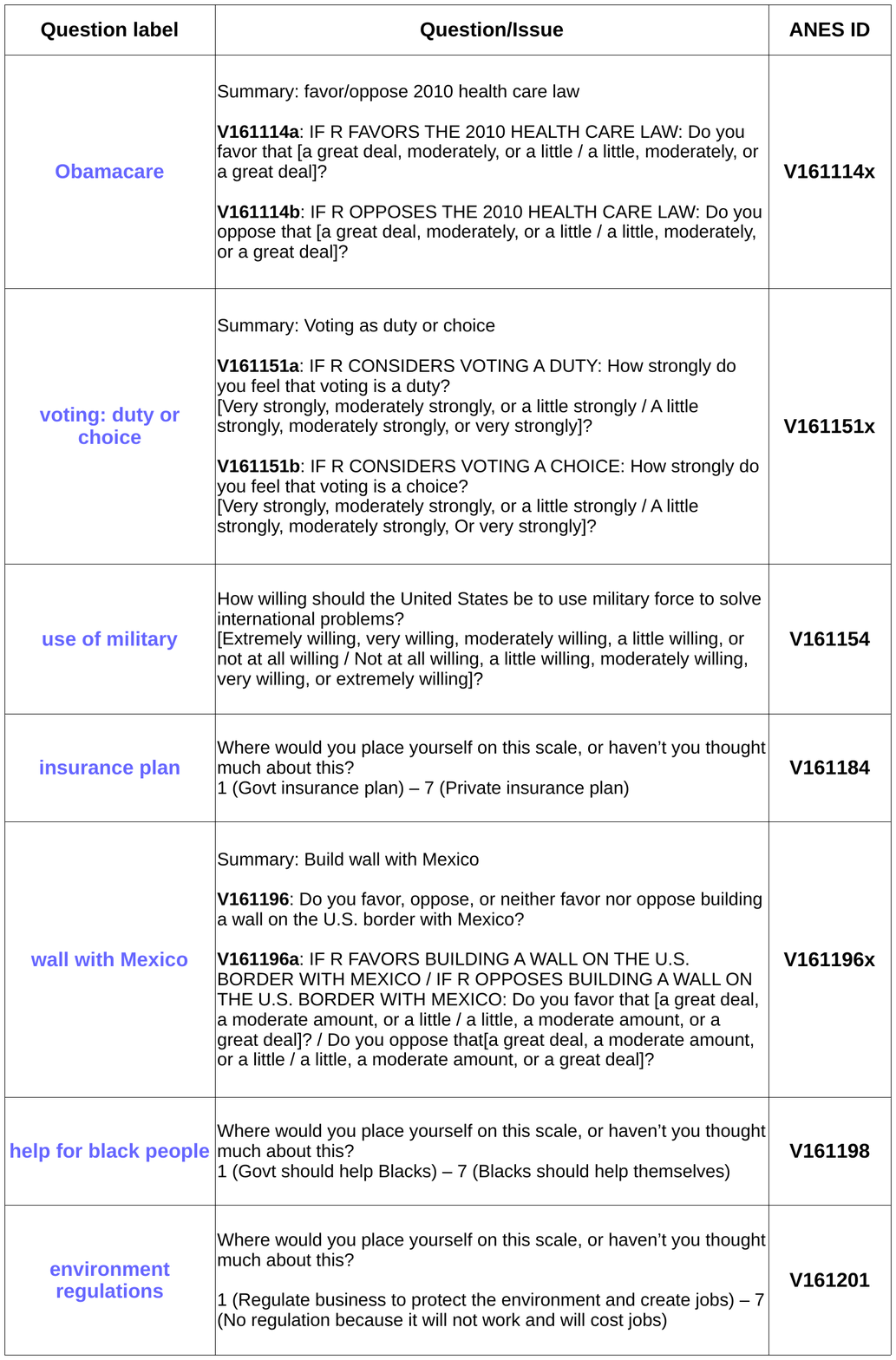}
\end{table}
\begin{table}[htb]
\includegraphics[width=\textwidth]{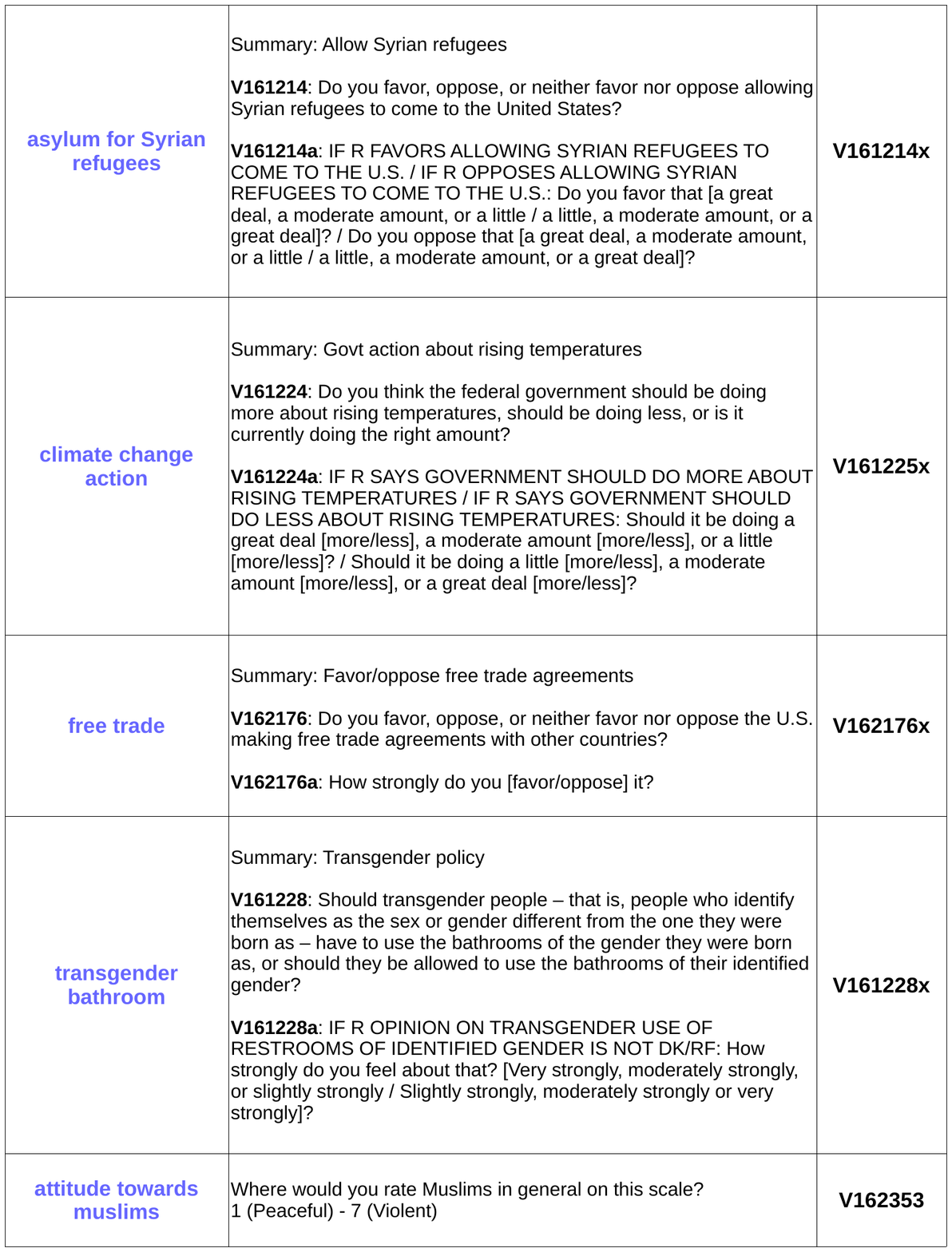}
\end{table}
\begin{table}[htb]
\includegraphics[width=\textwidth]{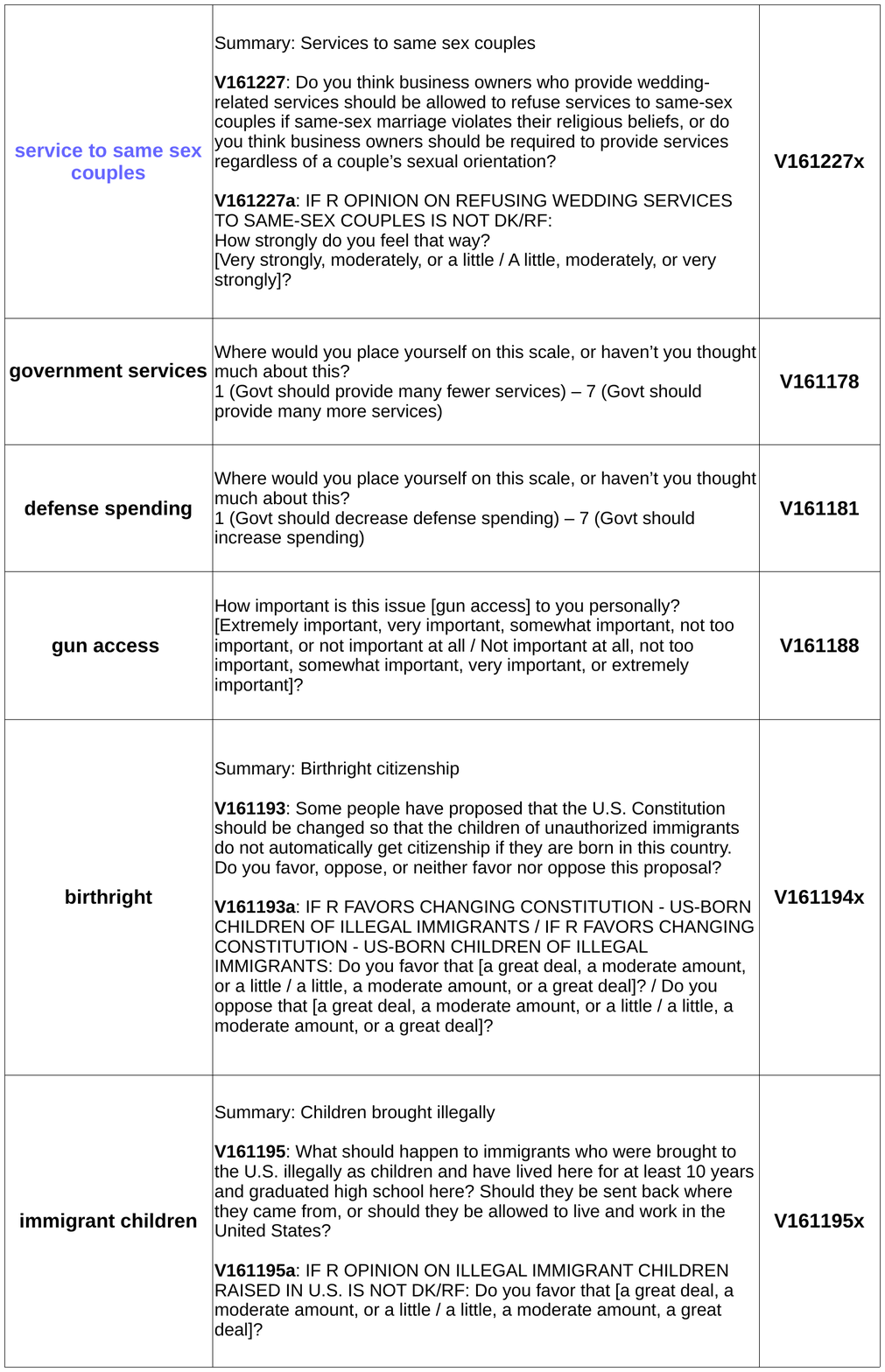}
\end{table}
\begin{table}[htb]
\includegraphics[width=\textwidth]{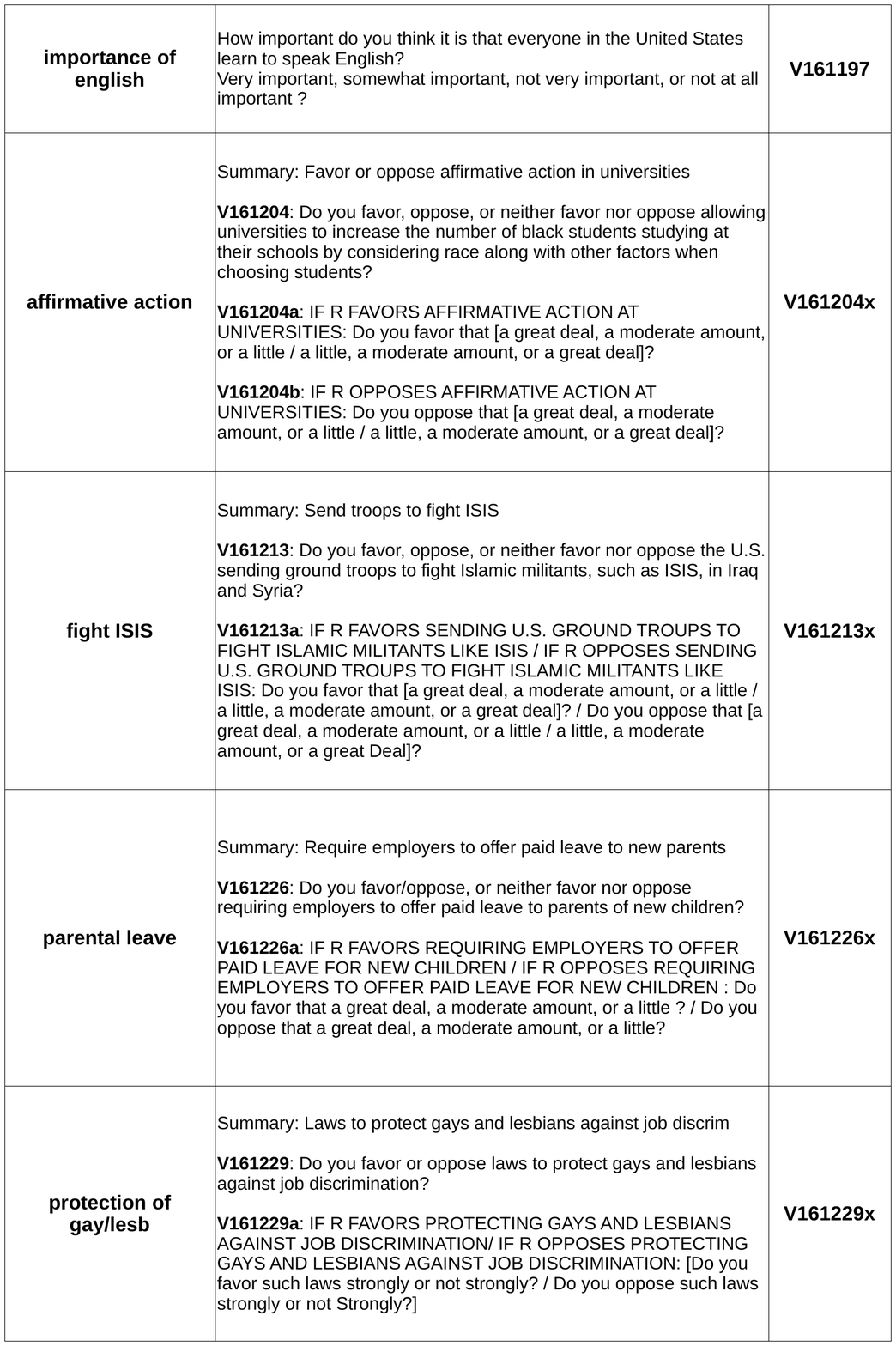}
\end{table}
\begin{table}[htb]
\includegraphics[width=\textwidth]{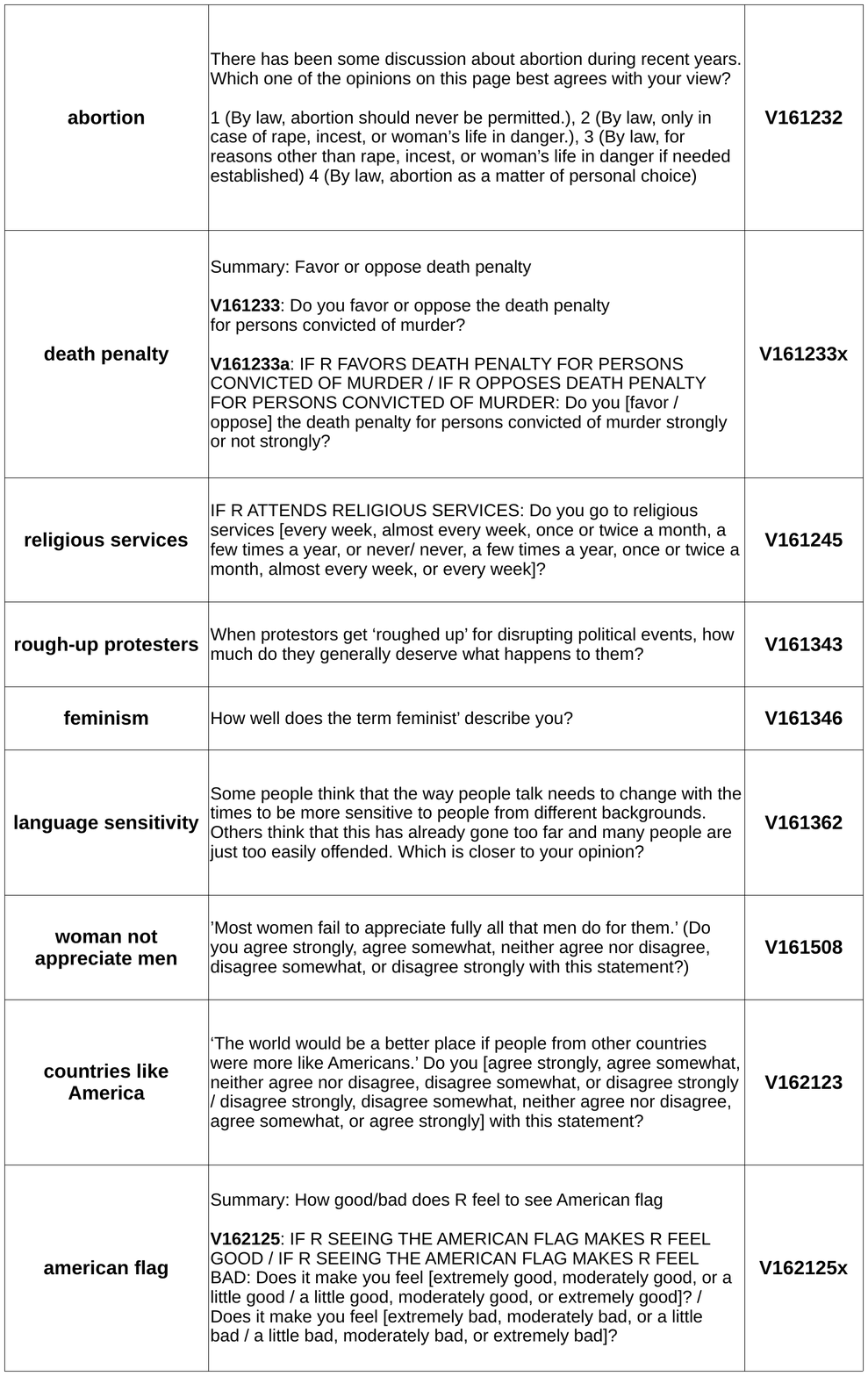}
\end{table}
\begin{table}[htb]
\includegraphics[width=\textwidth]{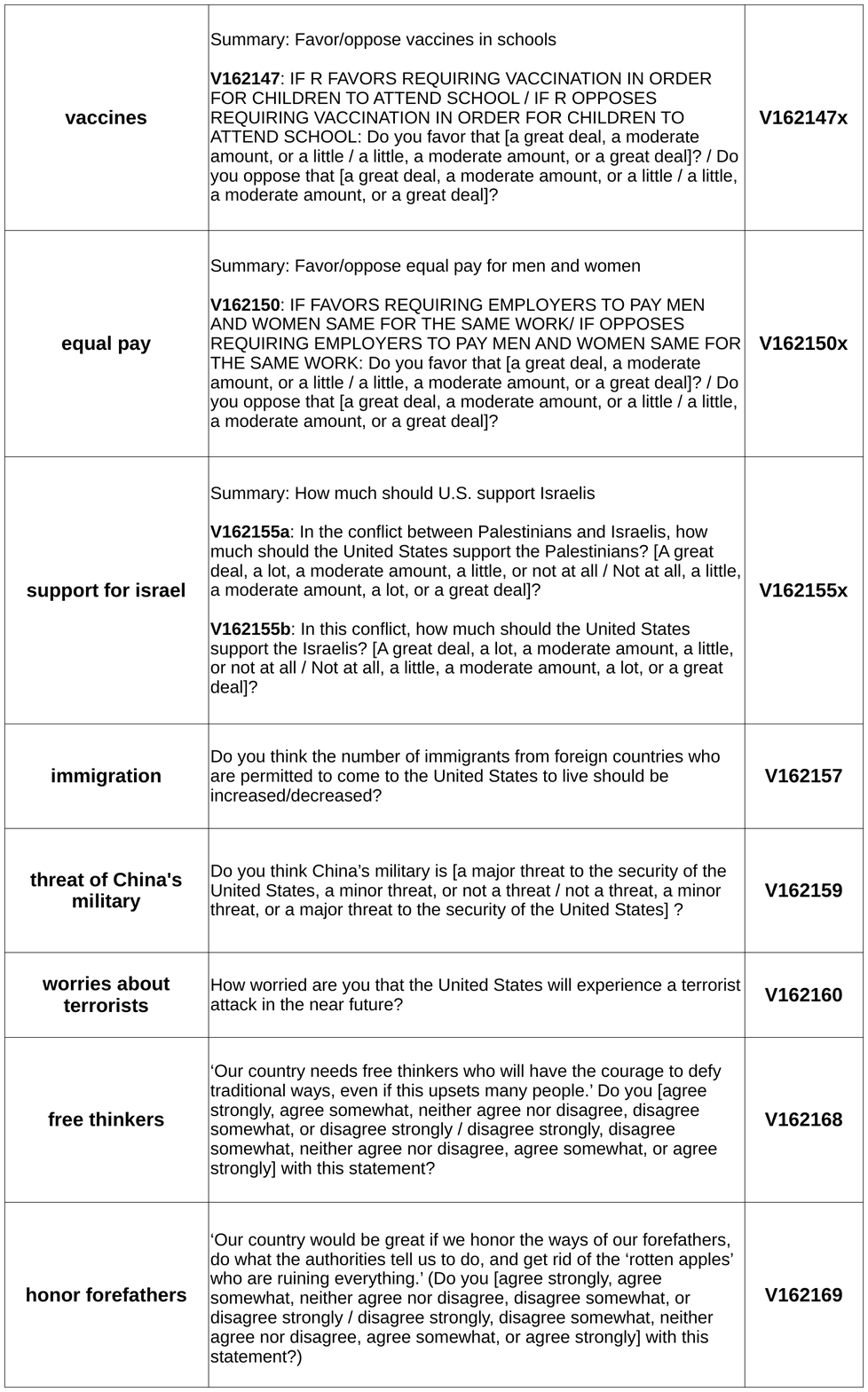}
\end{table}
\begin{table}[htb]
\includegraphics[width=\textwidth]{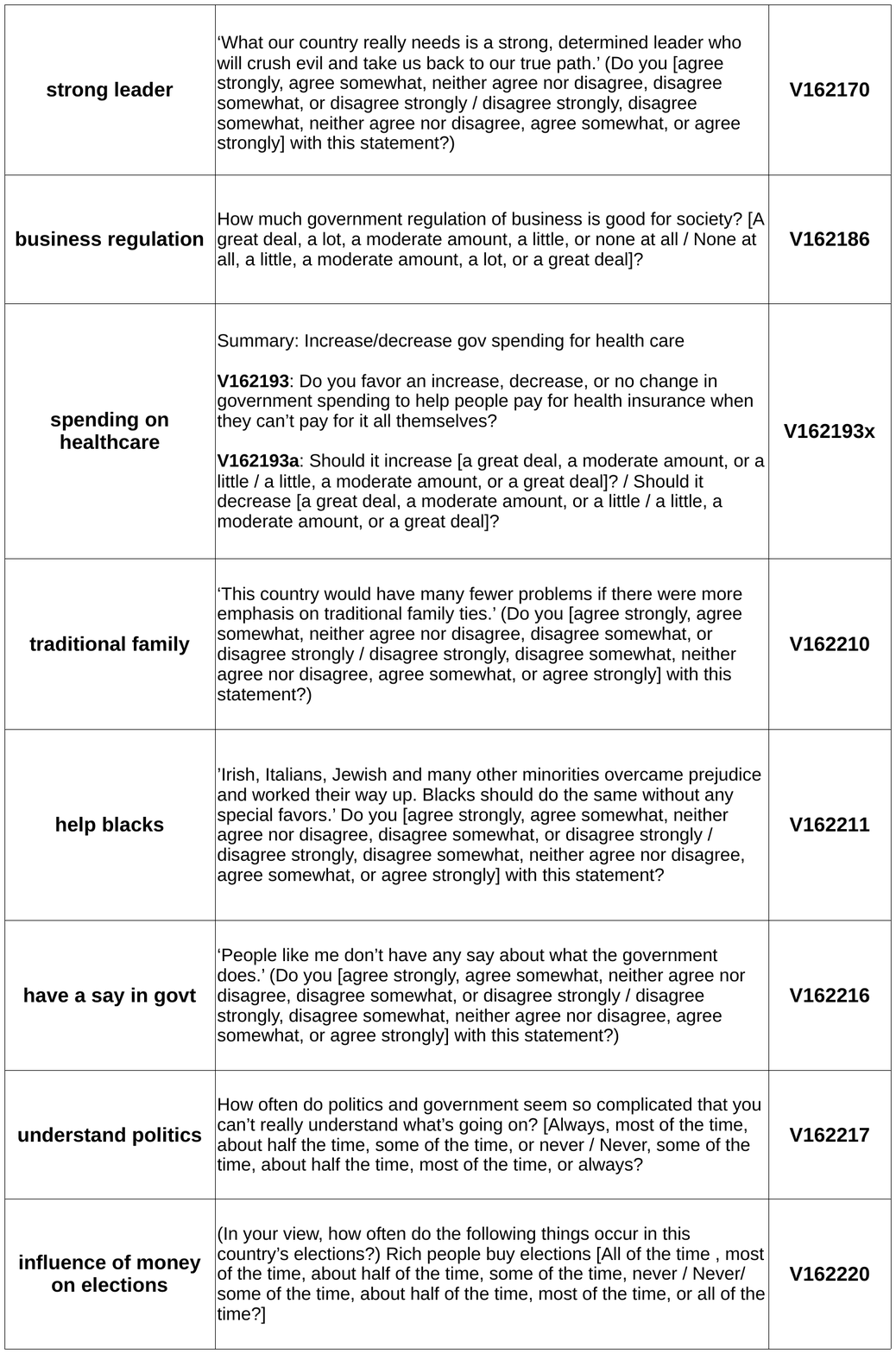}
\end{table}
\begin{table}[htb]
\includegraphics[width=\textwidth]{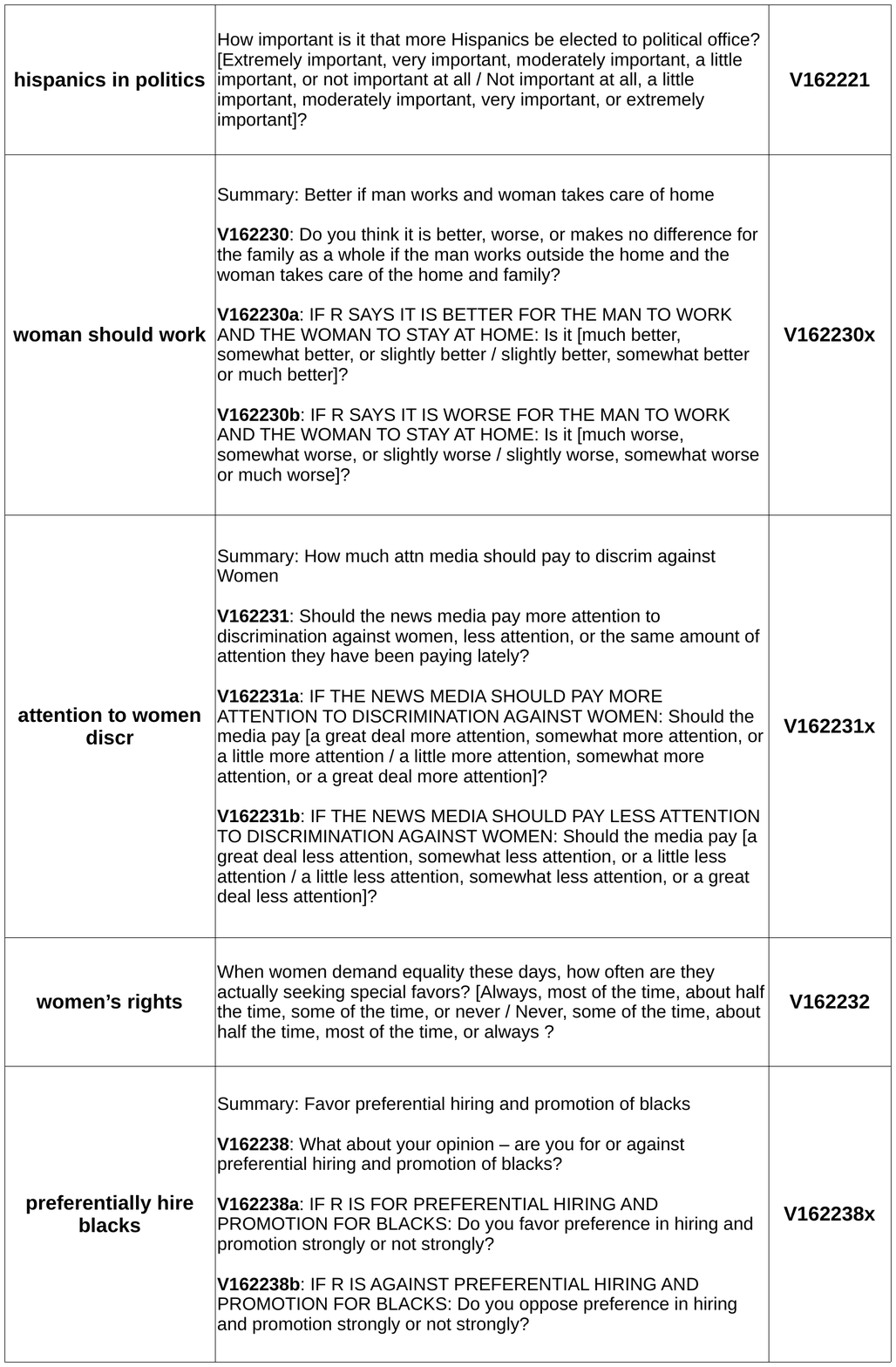}
\end{table}
\begin{table}[htb]
\includegraphics[width=\textwidth]{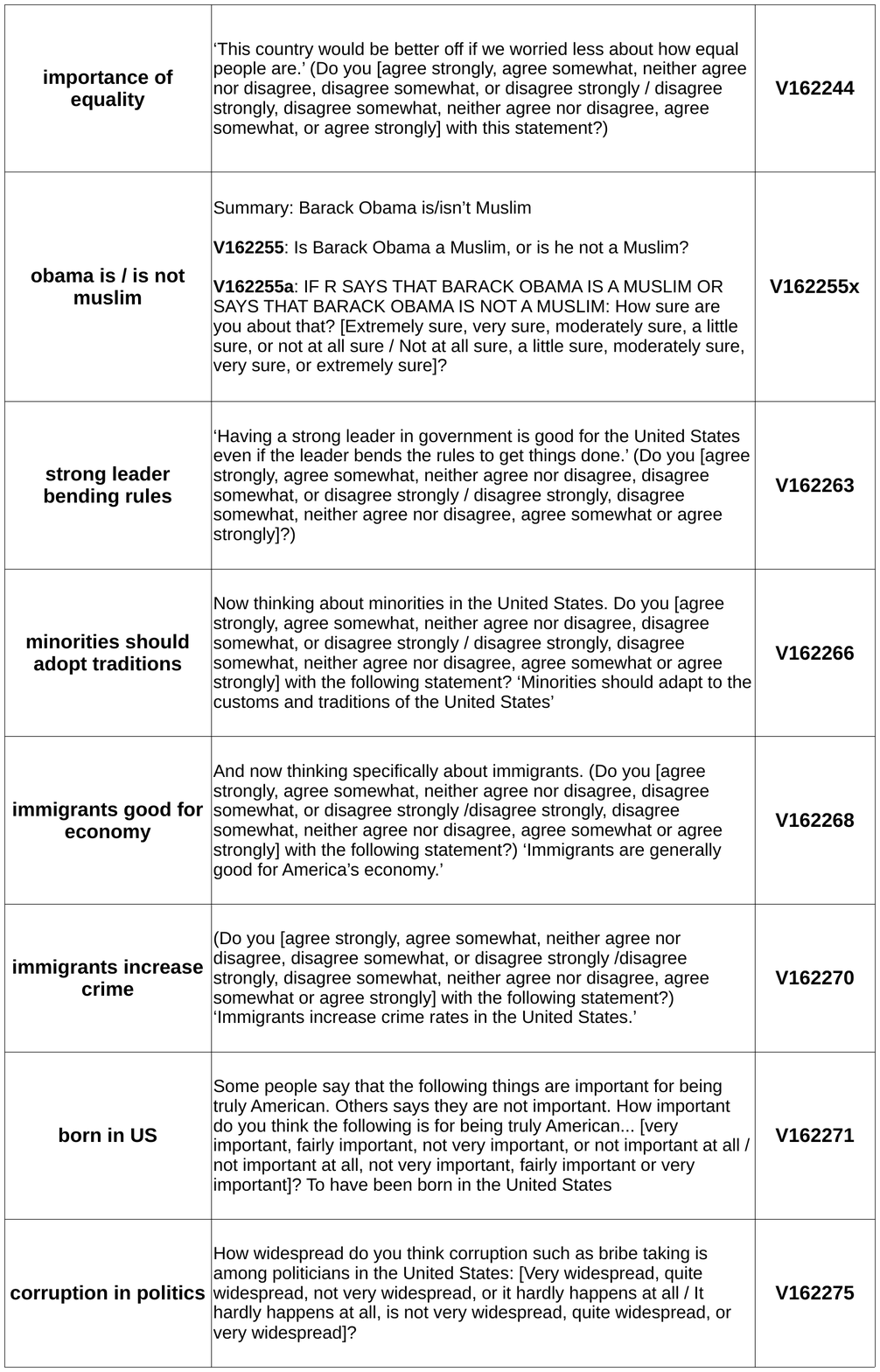}
\end{table}
\begin{table}[htb]
\includegraphics[width=\textwidth]{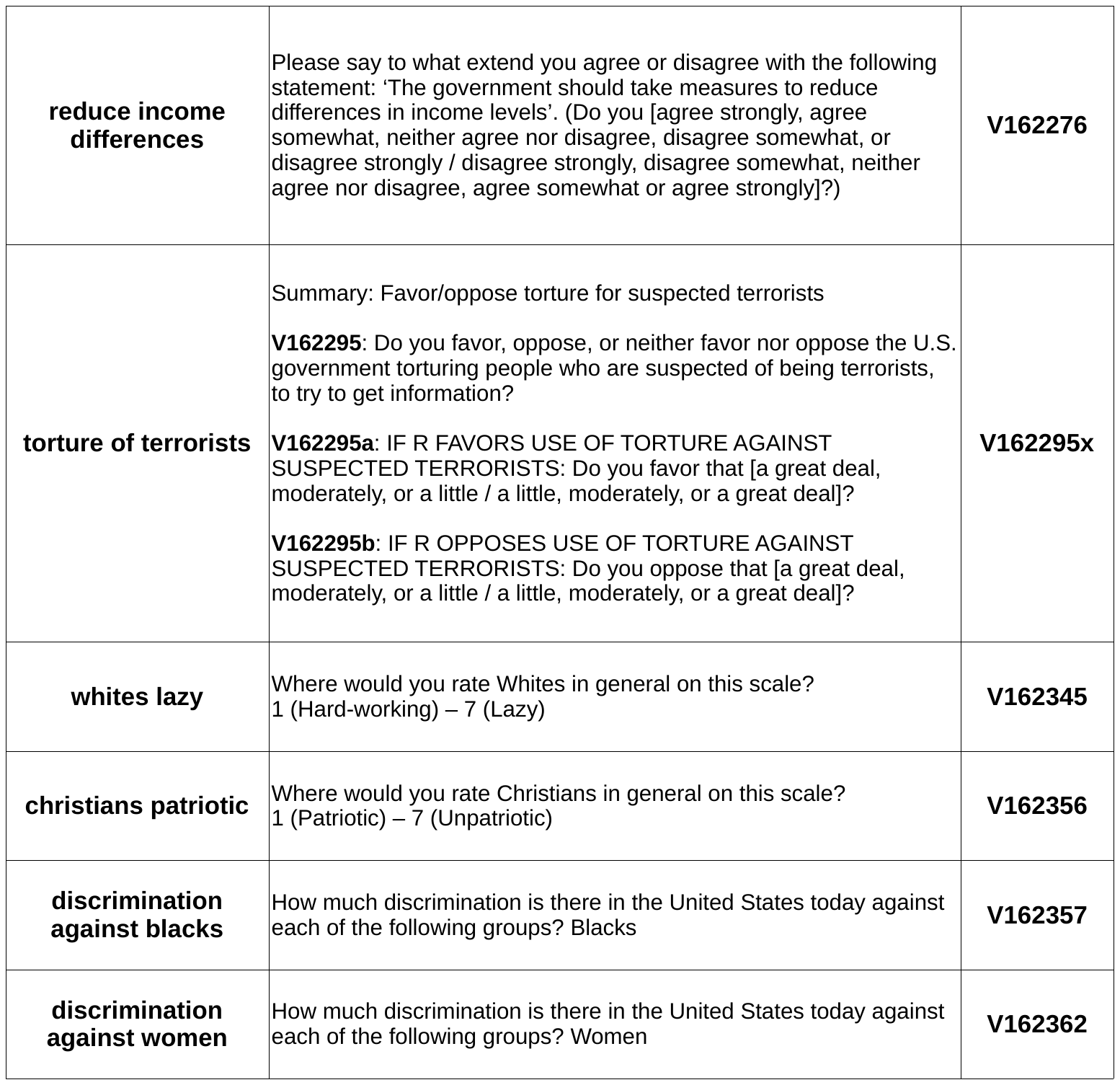}
\end{table}
\FloatBarrier
%\bibliographystyle{apsrev4-1}
%\bibliography{bib}

\end{document}